\documentclass[11pt,a4paper]{article}
\usepackage{graphicx}
\usepackage{amsmath}
\usepackage{amssymb}
\usepackage{cite}
\usepackage{tcolorbox}
\usepackage{subfigure}

\newtcolorbox{synthesis}{colback=white,colframe=blue!10!white,fonttitle=\bfseries,title=Synthesis,coltitle=black}
\newtcolorbox{outline}{colback=white,colframe=blue!10!white,fonttitle=\bfseries,title=Outline,coltitle=black}
\newtcolorbox{keywords}{colback=white,colframe=green!10!white,fonttitle=\bfseries,title=Keywords,coltitle=black}

\newcommand{\bm}{\boldsymbol}
\newcommand{\vect}[1]{\mathbf{#1}}
\newcommand{\mat}[1]{\mathbf{#1}}

\newcommand{\E}[1]{{\mathbb{E}}\{#1\}}

\newcommand{\etr}[1]{{\mathrm{etr}}\left\{#1\right\}}
\newcommand{\Tr}[1]{{\mathrm{Tr}}\{#1\}}

\renewcommand{\det}[1]{|#1|}

\newcommand{\pchol}[2]{\textrm{pchol}\left(#1,#2\right)}
\newcommand{\Proj}[1]{\mat{\Pi}_{#1}}
\newcommand{\Projorth}[1]{\mat{\Pi}^{\perp}_{#1}}

\newcommand{\rank}[1]{\mathrm{rank}\left(#1\right)}
\newcommand{\range}[1]{\mathcal{R}\left\{#1\right\}}

\newcommand{\sqnorm}[1]{\left\|#1\right\|^{2}}
\newcommand{\isqnorm}[1]{\left\|#1\right\|^{-2}}

\newcommand{\vbeta}{\bm{\beta}}

\newcommand{\mGamma}{\bm{\Gamma}}
\newcommand{\mLambda}{\bm{\Lambda}}
\newcommand{\mSigma}{\bm{\Sigma}}

\newcommand{\mOmega}{\bm{\Omega}}
\newcommand{\mOmegatilde}{\tilde{\mOmega}}
\newcommand{\mPsi}{\bm{\Psi}}

\newcommand{\efirst}{\vect{e}_{1}}
\newcommand{\etildefirst}{\tilde{\vect{e}}_{1}}

\newcommand{\vn}{\vect{n}}

\newcommand{\vq}{\vect{q}}

\newcommand{\vt}{\vect{t}}

\newcommand{\vtbar}{\bar{\vt}}
\newcommand{\vu}{\vect{u}}
\newcommand{\vv}{\vect{v}}
\newcommand{\vvn}{\vv_{n}}
\newcommand{\vvbar}{\bar{\vv}}
\newcommand{\vvtilde}{\tilde{\vv}}
\newcommand{\vw}{\vect{w}}

\newcommand{\vwtilde}{\tilde{\vw}}
\newcommand{\vwtildea}{\vwtilde_{a}}
\newcommand{\wcbf}{\vw_{\text{\tiny{wnmf}}}}
\newcommand{\wdl}{\vw_{\text{\tiny{dl}}}}
\newcommand{\wec}{\vw_{\text{\tiny{ec}}}}
\newcommand{\wecpchol}{\vw_{\text{\tiny{ec-pchol}}}}
\newcommand{\wmarzetta}{\vw_{\text{\tiny{Marzetta}}}}
\newcommand{\wmvdr}{\vw_{\text{\tiny{mvdr}}}}
\newcommand{\wmpdr}{\vw_{\text{\tiny{mpdr}}}}

\newcommand{\wopt}{\vw_{\text{\tiny{opt}}}}
\newcommand{\wtildeopt}{\vwtilde_{\text{\tiny{opt}}}}

\newcommand{\vx}{\vect{x}}
\newcommand{\vxbar}{\bar{\vx}}
\newcommand{\vxtilde}{\tilde{\vx}}

\newcommand{\vz}{\vect{z}}
\newcommand{\vztilde}{\tilde{\vz}}

\newcommand{\vzero}{\vect{0}}

\newcommand{\C}{\mSigma}
\newcommand{\Chat}{\hat{\C}}
\newcommand{\Ct}{\C_{t}}
\newcommand{\Ctilde}{\tilde{\C}}
\newcommand{\Cttilde}{\tilde{\Ct}}

\newcommand{\F}{\mat{F}}
\newcommand{\G}{\mat{G}}
\renewcommand{\H}{\mat{H}}
\newcommand{\I}{\mat{I}}
\newcommand{\eye}[1]{\I_{#1}}
\newcommand{\N}{\mat{N}}

\newcommand{\Q}{\mat{Q}}
\newcommand{\Qtilde}{\tilde{\Q}}

\renewcommand{\S}{\mat{S}}
\newcommand{\St}{\mat{S}_{t}}
\newcommand{\Sttilde}{\tilde{\mat{S}}_{t}}

\newcommand{\T}{\mat{T}}

\newcommand{\U}{\mat{U}}
\newcommand{\V}{\mat{V}}
\newcommand{\Vorth}{\V_{\perp}}

\newcommand{\W}{\mat{W}}
\newcommand{\Wtilde}{\tilde{\W}}

\newcommand{\X}{\mat{X}}
\newcommand{\Xt}{\X_{t}}
\newcommand{\Xttilde}{\tilde{\X}_{t}}
\newcommand{\Xbar}{\bar{\X}}
\newcommand{\Y}{\mat{Y}}

\newcommand{\Mzero}{\mat{0}}

\newcommand{\vCN}[3]{\mathbb{C}\mathcal{N}_{#1}\left(#2,#3\right)}
\newcommand{\mCN}[4]{\mathbb{C}\mathcal{N}_{#1}\left(#2,#3,#4\right)}

\newcommand{\mCT}[5]{\mathbb{C}\mathcal{T}_{#1}\left(#2,#3,#4,#5\right)}
\newcommand{\CW}[3]{\mathbb{C}\mathcal{W}_{#1}\left(#2,#3\right)}

\newcommand{\Cchisquare}[2]{\mathbb{C}\chi^{2}_{#1}(#2)}
\newcommand{\CF}[3]{\mathbb{C}\mathcal{F}_{#1}(#2,#3)}
\newcommand{\betapdf}[2]{\mathrm{Beta}(#1,#2)}
\newcommand{\dist}{\overset{d}{=}}
\newcommand{\approxdist}{\overset{d}{\approx}}

\newcommand{\alphaML}{\alpha_{\text{\tiny{ML}}}}
\newcommand{\onehalf}{\frac{1}{2}}

\newcommand{\SNR}{\mathrm{SNR}}

\newcommand{\SNRopt}{\SNR_{\text{\tiny{opt}}}}

\newcommand{\Beta}[2]{B_{#1,#2}}
\newcommand{\loss}{\rho}

\newcommand{\lossStudent}{\loss_{\text{\tiny{Student}}}}
\newcommand{\pdflossStudent}{p_{\text{\tiny{Student}}}}
\newcommand{\lossdl}{\loss_{\text{\tiny{dl}}}}
\newcommand{\lossec}{\loss_{\text{\tiny{ec}}}}
\newcommand{\lossecpchol}{\loss_{\text{\tiny{ec-pchol}}}}
\newcommand{\lossq}{\loss_{\C=\Ct+\vq\vq^{H}, \, \vq^{H}\Ct^{-1}\vv=0}}
\newcommand{\lossger}{\loss_{\text{\tiny{GER}}}}
\newcommand{\lossmpdr}{\loss_{\text{\tiny{mpdr}}}}
\newcommand{\pdflossmpdr}{p_{\text{\tiny{mpdr}}}}
\newcommand{\lossmvdr}{\loss_{\text{\tiny{mvdr}}}}
\newcommand{\pdflossmvdr}{p_{\text{\tiny{mvdr}}}}
\newcommand{\losspamvdr}{\loss_{\text{\tiny{pa-mvdr}}}}
\newcommand{\pdflosspamvdr}{p_{\text{\tiny{pa-mvdr}}}}
\newcommand{\losspampdr}{\loss_{\text{\tiny{pa-mpdr}}}}
\newcommand{\pdflosspampdr}{p_{\text{\tiny{pa-mpdr}}}}
\begin{document}
\title{A short overview of adaptive multichannel filters SNR loss analysis}
\author{Olivier Besson}
\date{September 2021}
\maketitle
\begin{outline}
Many multichannel systems use a linear filter to retrieve a signal of interest corrupted by noise whose statistics are partly unknown. The optimal filter in Gaussian noise requires knowledge of the noise covariance matrix $\C$ and in practice the latter is estimated from a set of training samples. An important issue concerns the characterization of the performance of such adaptive filters. This is generally achieved using as figure of merit the ratio of the signal to noise ratio (SNR) at the output of the adaptive filter to the SNR obtained with the clairvoyant -known $\C$- filter. This problem has been studied extensively since the seventies and this document presents a concise overview of results published in the literature. We consider various cases about the training samples covariance matrix and we investigate fully adaptive, partially adaptive and regularized filters. 
\end{outline}
\newpage

\section{Preamble}
This report focuses on performance analysis of multichannel adaptive filters through analysis of their SNR loss. It is an outgrowth of a course on array processing I have been giving for about fifteen years in various institutions in Toulouse at the Master of Sciences level. It should be viewed as a personal excerpt of the many results published in the literature since the seventies and it is in no way claimed to be exhaustive. Similarly it is not meant to provide an exhaustive list of references. Rather I will highlight along the document the references that have been most influential to me, starting with references \cite{VanTrees02,Kelly89,Ward94} by Van Trees, Kelly and Ward  which contain invaluable information. Very good overviews of topics related to this problem can also be found in \cite{Gershman03,Guerci03,Vorobyov14}.  The derivations leading to the representations of the SNR loss mostly borrow from results in multivariate statistics,  especially from the theory of Gaussian and Wishart distributions. Numerous references are available concerning this area, including \cite{Muirhead82,Gupta00,Goodman63,Khatri65}. 

\section{Problem statement}
Let us start with the problem of detecting the presence and/or estimating the  amplitude $\alpha$ of a known signal of interest (SoI) $\vv$ from a corrupted version $\vx=\alpha\vv+\vn$ where $\vn$ is the  disturbance and will be referred to as noise. In radar applications,  $\vv$ stands for the space and/or time signature of a potential target  and $\vn$ comprises clutter, thermal noise and possibly jammers.  The simplest processor is a linear filter $\vw$ which enables to estimate $\alpha$ as $\hat{\alpha}=\vw^{H}\vx$ and to decide of the SoI's presence, e.g. by comparing $|\vw^{H}\vx|^{2}$ to a threshold. The SNR at the output of the (fixed) filter $\vw$ is given by
\begin{equation}\label{def_SNR}
\SNR(\vw) = \frac{\E{|\alpha\vw^{H}\vv|^{2}}}{\E{|\vw^{H}\vn|^{2}}}	 = \frac{P|\vw^{H}\vv|^{2}}{\vw^{H}\C\vw}	
\end{equation}
When the noise $\vn$ follows a complex multivariate Gaussian distribution with zero mean and covariance matrix $\C$ the maximum likelihood estimate (MLE) of $\alpha$ writes $\alphaML = \wopt^{H} \vx$ with 
\begin{equation}\label{w_opt}
\wopt= \frac{\mSigma^{-1} \vv}{\vv^{H} \mSigma^{-1} \vv} 
\end{equation}
With white noise, i.e., when $\C=\gamma\eye{N}$, \eqref{w_opt} boils down to what we will refer to as the white noise matched filter $\wcbf=(\vv^{H}\vv)^{-1}\vv$.

The optimal filter $\wopt$ is also obtained as the solution to the following minimization problem:
\begin{equation}
\underset{\vw}{\min} \vw^{H}\mSigma\vw \text{ subject to } \vw^{H}\vv=1	
\end{equation}
In other words this filter minimizes the output power under the constraint that the signal of interest goes unscathed through the filter or equivalently $\wopt$ maximizes $\SNR(\vw)$.  Note that this interpretation holds irrespective of the distribution of $\vn$. $\wopt$ results in 
\begin{equation}\label{SNR_opt}
\SNRopt = \SNR(\wopt) = P \vv^{H}\C^{-1}\vv	
\end{equation}
 and consequently the performance of an arbitrary $\vw$ can be evaluated from the \emph{\textbf{SNR loss}}:
\begin{equation}\label{def_SNRloss}
\loss (\vw) = \frac{\SNR(\vw)}{\SNRopt}	 = \frac{|\vw^{H}\vv|^{2}}{(\vv^{H}\C^{-1}\vv)(\vw^{H}\C\vw)}	
\end{equation}
\section{Fully adaptive processing}
Let us consider the practical situation where $\C$ is unknown and hence needs to be inferred from a set of $K$ samples which are usually referred to as training samples.  We assume in this section that $K \geq N$. Let us assume that the training samples are independent and identically distributed according to a multivariate Gaussian distribution with zero-mean and covariance matrix $\Ct$ which, at this stage, is arbitrary and possibly different from $\C$. 

Let $\Xt$ be the $N|K$ matrix gathering the training samples so that, with the notations described in the appendix,  $\Xt \dist \mCN{N,K}{\Mzero}{\Ct}{\eye{K}}$. Let $\St = \Xt \Xt^{H}$ denote  the sample covariance matrix and let us build the adaptive matched filter  
\begin{equation}\label{w_fa}
\vw = (\vv^{H} \St^{-1} \vv)^{-1} \St^{-1} \vv	
\end{equation}
In order to evaluate its performance we will use the SNR loss which, with the above choice of $\vw$, becomes
\begin{equation}\label{SNRloss}
\loss = \frac{(\vv^H \St^{-1} \vv)^{2}}{(\vv^{H}\mSigma^{-1}\vv)(\vv^{H}\St^{-1}\mSigma\St^{-1}\vv)}
\end{equation}
Since $\St$ is random $\loss$ is a random variable whose probability density function (p.d.f.) we are interested in. 

Before continuing, we would like to make the following comments. First note that the case where \textcolor{blue}{$\Ct=\C$} is of primary importance and has been studied in the fundamental work by Reed, Mallett and Brenann  \cite{Reed74} with a radar application point of view, see also \cite{Hanumara86,Khatri87} for a multivariate analysis oriented study. Following Van Trees's terminology \cite{VanTrees02}, we will refer to this case as the minimum variance distortionless response (\textcolor{blue}{MVDR}) scenario.  Now in some applications there is possibly a covariance mismatch between the samples that are used to train the filter and the samples to be filtered, and a large  number of papers have addressed this issue. The simplest case is the so-called partially homogeneous environment where $\Ct = \gamma \C$ \cite{Kraut99,Kraut05}. A second common example is the case  where the training samples contain the SoI, i.e., \textcolor{cyan}{$\Ct=\C+P\vv\vv^{H}$}, a scenario which will be referred to as the minimum power distortionless response (\textcolor{cyan}{MPDR}) scenario. Its thorough analysis can be found in \cite{Boroson80}. The training samples can also be contaminated by signal-like components or outliers \cite{Gerlach95,Gerlach02} or there might exist a  rank-one difference between $\Ct$ and $\C$, for instance a surprise or undernulled interference  \cite{Richmond00c,Besson07c,Besson07g}.  The case where $\Ct$ and $\C$ are different but satisfy the so-called generalized eigen-relation is dealt with in \cite{Richmond00b,Blum00} while an arbitrary $\Ct$ is considered in \cite{McDonald00,Raghavan19,Besson20g,Besson21}.

In what follows we let $\Ct^{1/2}$ denote a square-root of $\Ct$ i.e., $\Ct = \Ct^{1/2} (\Ct^{1/2})^{H}$ and we use the short-hand notations $\Ct^{H/2}=(\Ct^{1/2})^{H}$, $\Ct^{-1/2}=(\Ct^{1/2})^{-1}$ and $\Ct^{-H/2}=(\Ct^{H/2})^{-1}$. The definition of the complex matrix-variate distributions appearing below can be found in the appendix.
\vspace*{0.2cm}
\subsection{Analysis of the SNR loss}
We consider first an arbitrary $\Ct$ and then specialize to the above cases of interest. Our aim is to obtain a statistical representation of $\loss$ in terms of well-known distributions and the derivations below follow from \cite{Besson20g}. Since $\Xt \dist \Ct^{1/2} \N$ with $\N \dist \mCN{N,K}{\Mzero}{\eye{N}}{\eye{K}}$, it ensues that $\St \dist \Ct^{1/2} \W \Ct^{H/2}$ with $\W \dist \CW{N}{K}{\eye{N}}$. Therefore, one can write
\begin{align}\label{deriv_rep_rho}
\loss &= \frac{(\vv^H \St^{-1} \vv)^{2}}{(\vv^{H}\C^{-1}\vv)(\vv^{H}\St^{-1}\C\St^{-1}\vv)} \nonumber \\
&\dist  \frac{(\vv^H \Ct^{-H/2} \W^{-1} \Ct^{-1/2}  \vv)^{2}}{(\vv^{H}\C^{-1}\vv)(\vv^{H}\Ct^{-H/2}  \W^{-1}  \Ct^{-1/2} \C \Ct^{-H/2} \W^{-1}  \Ct^{-1/2} \vv)} \nonumber \\
&\dist  \frac{(\vv^H \Ct^{-H/2} \Q \W^{-1} \Q^{H} \Ct^{-1/2}  \vv)^{2}}{(\vv^{H}\C^{-1}\vv)(\vv^{H}\Ct^{-H/2} \Q  \W^{-1}  \Q^{H} \Ct^{-1/2} \C \Ct^{-H/2} \Q \W^{-1} \Q^{H}  \Ct^{-1/2} \vv)}
\end{align}
for any unitary matrix $\Q$ since $\W$ and $\Q\W\Q^{H}$ have the same distribution. Let us define
\begin{align}\label{Omega}
\mOmega &= \Q^{H} \Ct^{-1/2} \C \Ct^{-H/2} \Q  = \begin{pmatrix} \Omega_{11} & \mOmega_{12} \\ \mOmega_{21} & \mOmega_{22} \end{pmatrix}
\end{align}
and let us choose $\Q$ such that $\Q^{H}  \Ct^{-1/2} \vv = (\vv^{H}\Ct^{-1}\vv)^{1/2} \efirst$ where $\efirst = \begin{bmatrix} 1 & 0 & \ldots & 0 \end{bmatrix}^{T}$. With this choice we obtain
\begin{equation}
\loss 	\dist  \frac{\vv^{H}\Ct^{-1}\vv}{\vv^{H}\C^{-1}\vv} \; \frac{(\efirst^{H}\W^{-1}\efirst)^{2}}{\efirst^{H}\W^{-1}\mOmega \W^{-1}\efirst}
\end{equation}
If we partition $\W$ as in \eqref{Omega} we get
\begin{align}
\W^{-1}\efirst &=  \begin{pmatrix} W_{1.2}^{-1} & -W_{1.2}^{-1} \W_{12 } \W_{22}^{-1} \\ - \W_{22}^{-1} \W_{21} W_{1.2}^{-1} & \W_{2.1}^{-1} \end{pmatrix}  \begin{pmatrix} 1 \\ \vzero  \end{pmatrix} \nonumber \\
& = W_{1.2}^{-1} \begin{pmatrix}  1 \\ -\vt_{21 }  \end{pmatrix}
\end{align}
where 
\begin{equation}
W_{1.2}=W_{11} - \W_{12}\W_{22}^{-1} \W_{21 }; \quad \vt_{21} = \W_{22}^{-1} \W_{21}
\end{equation}
The SNR loss can thus be written as
\begin{equation}
\loss = \frac{\vv^{H}\Ct^{-1}\vv}{\vv^{H}\C^{-1}\vv}  \; \frac{1}{\begin{bmatrix}  1 &-\vt_{21}^{H}  \end{bmatrix} \mOmega \begin{bmatrix} 1 \\ -\vt_{21} \end{bmatrix}  }
\end{equation}
Moreover
\begin{align}
\begin{bmatrix}  1 &-\vt_{21}^{H}  \end{bmatrix} \mOmega \begin{bmatrix} 1 \\ -\vt_{21} \end{bmatrix}   &= \Omega_{11} - \vt_{21}^{H} \mOmega_{21}  - \mOmega_{12}\vt_{21} + \vt_{21}^{H} \mOmega_{22} \vt_{21} \nonumber \\
&=  \Omega_{1.2} + (\vt_{21}-\mOmega_{22}^{-1}\mOmega_{21})^{H} \mOmega_{22} (\vt_{21}-\mOmega_{22}^{-1}\mOmega_{21})
\end{align}
which, along with the readily verified fact that $\Omega_{1.2} =  \frac{\vv^{H}\Ct^{-1}\vv}{\vv^{H}\C^{-1}\vv}$ yields the following compact expression
\begin{equation}\label{SNRloss_vs_Omega}
\loss = \left[ 1+ \frac{\vv^{H}\C^{-1}\vv}{\vv^{H}\Ct^{-1}\vv} (\vt_{21}-\vtbar_{21})^{H} \mOmega_{22} (\vt_{21}-\vtbar_{21}) \right]^{-1}
\end{equation}
with $\vtbar_{21}=\mOmega_{22}^{-1}\mOmega_{21}$. From \cite{Khatri87} $\vt_{21}$ follows a complex multivariate Student distribution and can be represented as
\begin{equation}
\vt_{21} \dist \frac{\vn_{21}}{\sqrt{V_{21}}} \dist \frac{\vCN{N-1}{\vzero}{\eye{N-1}}}{\sqrt{\Cchisquare{K-N+2}{0}}}
\end{equation}
 Let $\mOmega_{22} = \U \mLambda \U^{H}$ denote the eigenvalue decomposition of $\mOmega_{22}$. Then one can write the quadratic form in \eqref{SNRloss_vs_Omega} as
\begin{align}\label{rep_Q}
Q &= (\vt_{21}-\vtbar_{21})^{H} \mOmega_{22} (\vt_{21}-\vtbar_{21}) \nonumber \\
&= V_{21}^{-1} (\vn_{21}-V_{21}^{1/2}\vtbar_{21})^{H} \U \mLambda \U^{H} (\vn_{21}-V_{21}^{1/2}\vtbar_{21})  \nonumber \\
&= V_{21}^{-1} (\U^{H}\vn_{21}-V_{21}^{1/2}\U^{H}\vtbar_{21})^{H} \mLambda  (\U^{H}\vn_{21}-V_{21}^{1/2}\U^{H}\vtbar_{21})  \nonumber \\
&= V_{21}^{-1} \sum_{i=1}^{N-1} \lambda_{i} \left| \vu_{i}^{H}(\vn_{21}-V_{21}^{1/2}\vtbar_{21})\right|^{2} \nonumber \\
&\dist V_{21}^{-1} \sum_{i=1}^{N-1} \lambda_{i} \Cchisquare{1}{V_{21} |\vu_{i}^{H}\vtbar_{21}|^{2}} \nonumber \\
&\dist V_{21}^{-1} \sum_{i=1}^{N-1} \lambda_{i} \Cchisquare{1}{V_{21} \delta_{i}}
\end{align}
where $V_{21} \dist \Cchisquare{K-N+2}{0}$ and $\delta_{i}=| \vu_{i}^{H}\vtbar_{21}|^{2}$. Therefore, the SNR loss admits the following representation:
\begin{equation}\label{rep_SNRloss_general}
\boxed{\loss \dist \left[ 1+ \frac{\vv^{H}\C^{-1}\vv}{\vv^{H}\Ct^{-1}\vv} \; \frac{\sum_{i=1}^{N-1} \lambda_{i} \Cchisquare{1}{V_{21}  \delta_{i}}}{V_{21}} \right]^{-1}}
\end{equation}
\textbf{\emph{The previous equation provides a statistical representation of the SNR loss for arbitrary matrices $\C$ and $\Ct$}}. 

\begin{figure}[h]
\centering
\includegraphics[width=11cm]{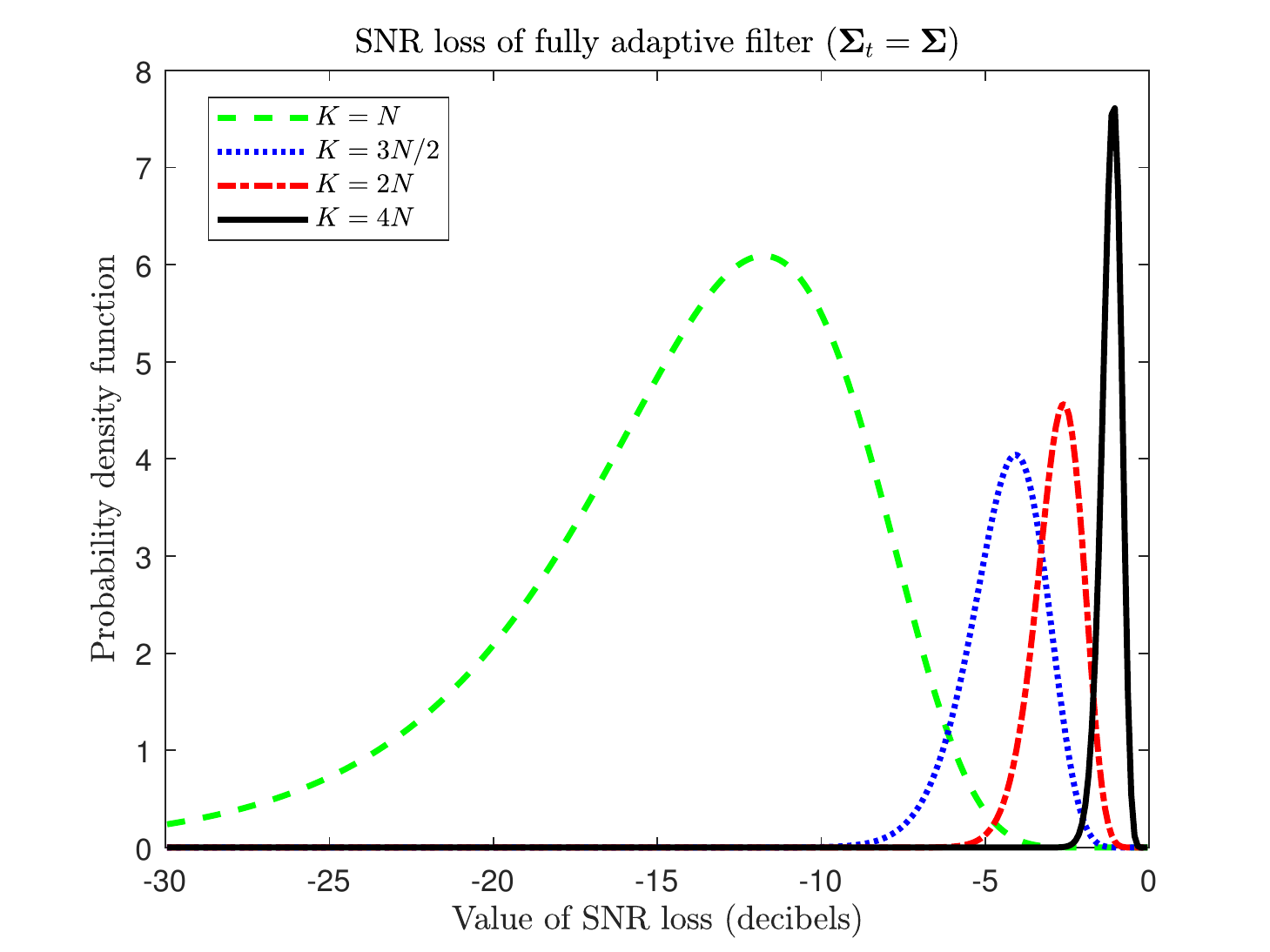}
\caption{Distribution of the SNR loss when $\Ct=\C$. $N=16$ and varying $K$.}
\label{fig:pdf_snrloss_fa_mvdr}
\end{figure}

We will examine later on the impact of $\Ct \neq \C$ on the SNR loss distribution. Prior to that, let us consider  the MVDR scenario for which $\Ct=\C$ or possibly $\Ct = \gamma \C$ since it does not change the distribution of the SNR loss. In this case, one has $\mOmega=\gamma^{-1}\eye{N}$, $\lambda_{i}=\gamma^{-1}$ and $\delta_{i}=0$, which results in
\begin{equation}\label{rep_SNRloss_MVDR}
\boxed{\lossmvdr \dist \left[1 + \frac{\Cchisquare{N-1}{0}}{\Cchisquare{K-N+2}{0}}\right]^{-1} }
\end{equation}
and therefore $\lossmvdr \dist \betapdf{N-1}{K-N+2}$ with a p.d.f. given by
\begin{equation}\label{pdf_SNRloss_MVDR}
\pdflossmvdr(\loss) = 	\frac{1}{\Beta{N-1}{K-N+2}} \loss^{K-N+1} (1-\loss)^{N-2}
\end{equation}
\emph{The distribution of $\lossmvdr$ is therefore independent of $\C$ and depends only on $N$ and $K$}. It is straightforward to see that $\E{\lossmvdr}=(K-N+2)/(K+1)$ so that
\begin{equation}\label{RMB_rule}
\E{\lossmvdr}=0.5 \Leftrightarrow K=2N-3
\end{equation} 
which corresponds to the famous \textbf{Reed-Mallet-Brennan rule} \cite{Reed74}. As an illustration, we display in Figure \ref{fig:pdf_snrloss_fa_mvdr} the Beta distribution \eqref{pdf_SNRloss_MVDR} for different values of $K$ with $N=16$. As can be seen the fully adaptive processor does a good job only when a sufficient number of training samples is available whereas when $K$ approaches $N$ the support of this distribution is significantly moved towards very small values, and one may wonder about using $\vw = (\vv^{H} \St^{-1} \vv)^{-1} \St^{-1} \vv$ under these circumstances. Note that the requirement for a large number of training samples can somehow be relieved if one assumes some structure for $\C$. For instance if $\C$ is known to be persymmetric and this property is exploited then the corresponding fully adaptive filter has a $\betapdf{\frac{N-1}{2}}{\frac{2K-N+2}{2}}$  distribution \cite{Liu16} and hence approximately $N$ samples are required to achieve convergence.

\begin{figure}[h]
\centering
\includegraphics[width=11cm]{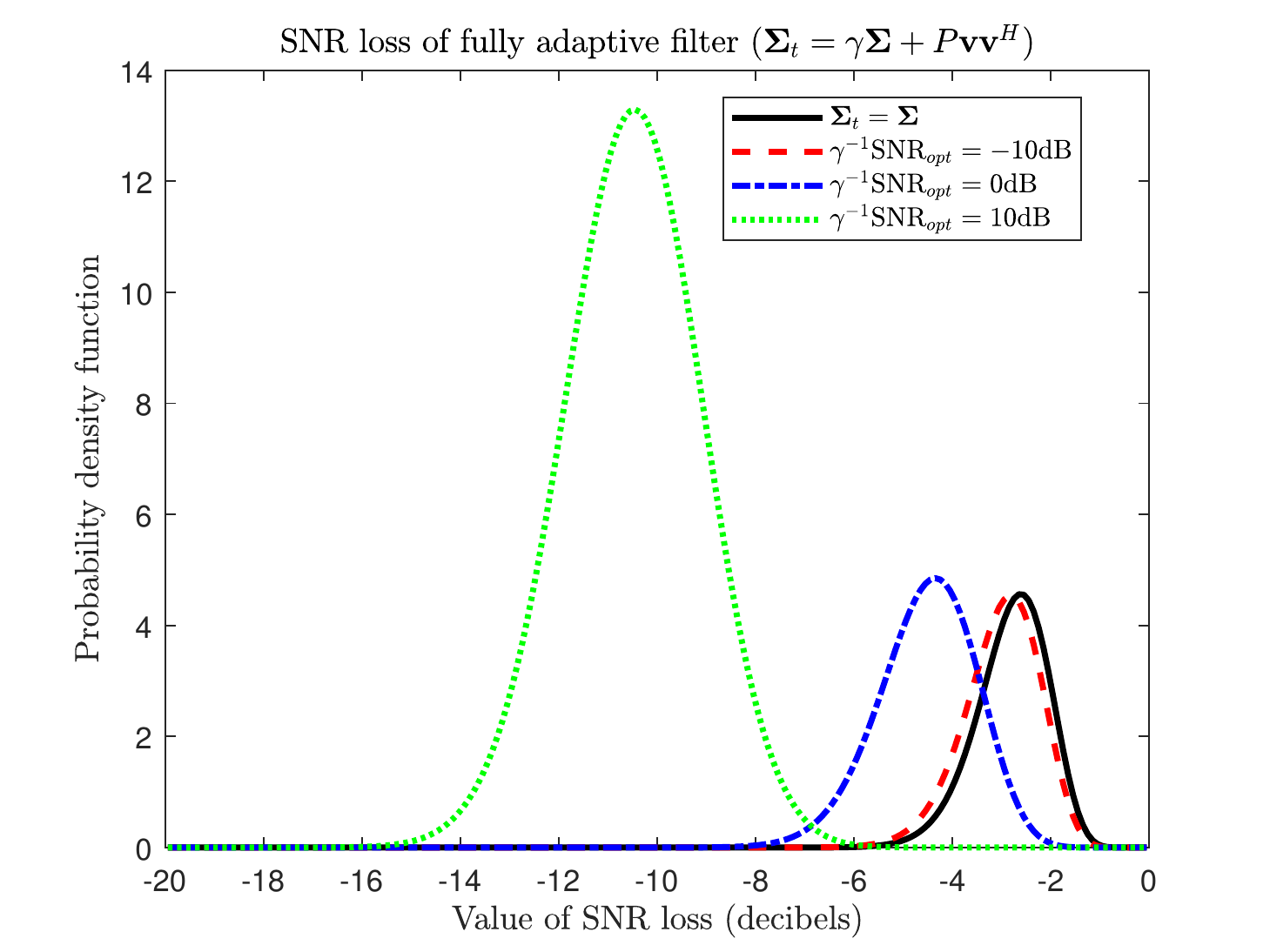}
\caption{Distribution of the SNR loss in the presence of the SoI in the training samples versus $\gamma^{-1}\SNRopt$. $N=16$ and $K=2N$.}
\label{fig:pdf_sinrloss_MPDR_K=32}
\end{figure}

Let us now move to the MPDR scenario  which has been thoroughly studied by Boroson \cite{Boroson80}. Actually the latter reference considers a rather large set of scenarios (MVDR or MPDR) and additionally takes into account errors about the SoI signature, i.e., Boroson derives representations of the SNR loss of $\vw \propto \St^{-1} \vvbar$ where $\vvbar \neq \vv$ for both the MVDR and MPDR scenarios. Assuming no SoI signature error and a partially homogeneous MPDR scenario -$\Ct = \gamma \C + P \vv \vv^{H}$- it can be shown \cite{Besson20g} that
\begin{equation}\label{rep_SNRloss_MPDR}
\boxed{\lossmpdr \dist \left[1 + (1+\gamma^{-1}\SNRopt) \frac{\Cchisquare{N-1}{0}}{\Cchisquare{K-N+2}{0}}\right]^{-1}}
\end{equation}
which \emph{depends only on $K$, $N$ and $\gamma^{-1}\SNRopt$}. It is clear that the difference between the MVDR and the MPDR scenarios will be all the more pronounced that $\gamma^{-1}\SNRopt$ is large. The  p.d.f. of $\lossmpdr$ is given by
\begin{align}\label{pdf_SNRloss_MPDR}
\pdflossmpdr(\loss) &= \frac{\loss^{K-N+1} (1-\loss)^{N-2}}{\Beta{N-1}{K-N+2}} \frac{(1+\gamma^{-1}\SNRopt)^{K-N+2}}{(1+\loss\gamma^{-1}\SNRopt)^{K+1}} \nonumber \\
&= \pdflossmvdr(\loss) \times \frac{(1+\gamma^{-1}\SNRopt)^{K-N+2}}{(1+\loss\gamma^{-1}\SNRopt)^{K+1}}
\end{align}
The detrimental effect of the presence of the SoI in the training samples is illustrated in Figure \ref{fig:pdf_sinrloss_MPDR_K=32} where we plot the distribution of $\lossmpdr$ for $K=2N$ and various values of $\gamma^{-1}\SNRopt$ and where we compare it with the distribution of $\lossmvdr$. Obviously when $\gamma^{-1}\SNRopt$ increases a large difference is observed and $\lossmpdr$ is likely to take very low values which makes the interest of the adaptive filter questionable in this situation.

\begin{figure}[h]
	\centering
	\includegraphics[width=11cm]{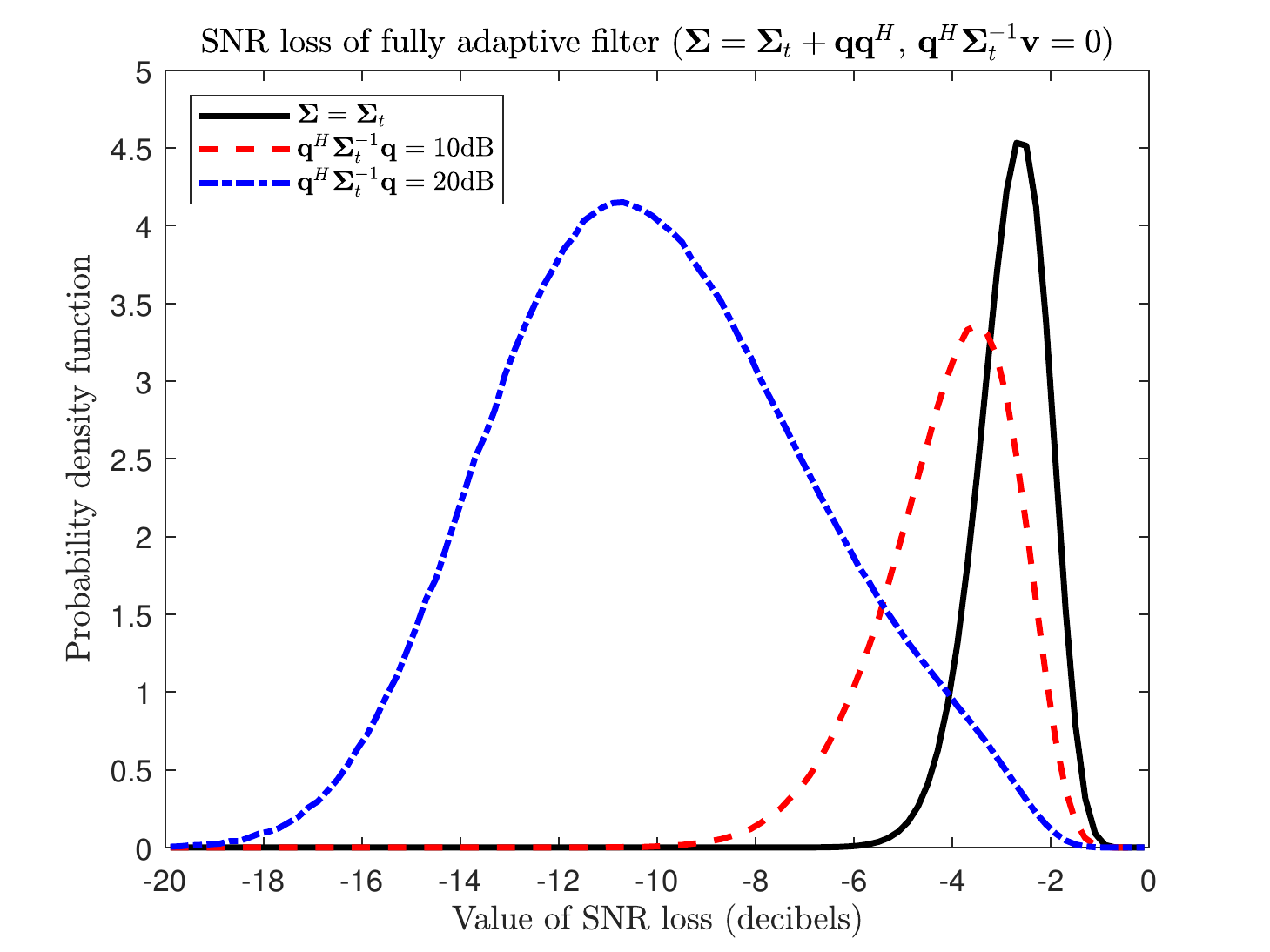}
	\caption{Distribution of the SNR loss when $\C=\Ct+\vq\vq^{H}$ and $\vq^{H}\Ct^{-1}\vv=0$. $N=16$ and $K=2N$.}
	\label{fig:pdf_snrloss_ger_undernulled_K=32}
\end{figure}

Another case of interest stems from observing the importance of the vector $\vtbar_{21}$ in \eqref{rep_Q}. Indeed the non-centrality parameter $\delta_{i}$ of \eqref{rep_SNRloss_general} depends on it and therefore becomes $0$ if $\vtbar_{21}=\vzero$. The latter condition is equivalent to the so-called \textbf{\emph{generalized eigen-relation}} (GER) \cite{Richmond00c,Richmond00b} which states that $\Ct^{-1}\vv = \lambda \C^{-1}\vv$. If the latter condition is fulfilled then
\begin{equation}\label{rep_SNRloss_GER}
\boxed{\lossger \dist \left[ 1+  \frac{\sum_{i=1}^{N-1} \lambda_{i} \Cchisquare{1}{0}}{\lambda\Cchisquare{K-N+2}{0}} \right]^{-1}}
\end{equation}
Before illustrating the general case, we investigate a special case of the GER, namely when $\C=\Ct+\vq\vq^{H}$ and $\vq^{H}\Ct^{-1}\vv=0$, i.e., the data to be filtered contains a rank-one component (e.g. a surprise interference) which is not accounted for by the training samples and which falls in a null of the optimal filter $\wopt$. In this case, the expression in \eqref{rep_SNRloss_GER} can be simplified to
\begin{equation}\label{rep_SNRloss_undernulled}
\lossq \dist \left[1 + \frac{\Cchisquare{N-2}{0} + (1+\vq^{H}\Ct^{-1}\vq) \Cchisquare{1}{0} }{\Cchisquare{K-N+2}{0}}\right]^{-1}
\end{equation}
which depends on $\vq^{H}\Ct^{-1}\vq$ only. In Figure  \ref{fig:pdf_snrloss_ger_undernulled_K=32} we illustrate the impact of such a surprise interference which is present in the data to be filtered but has not been learned from the training samples. With the power of this interference increasing the degradation is seen to be substantial.

\begin{figure}[h]
	\centering
	\includegraphics[width=11cm]{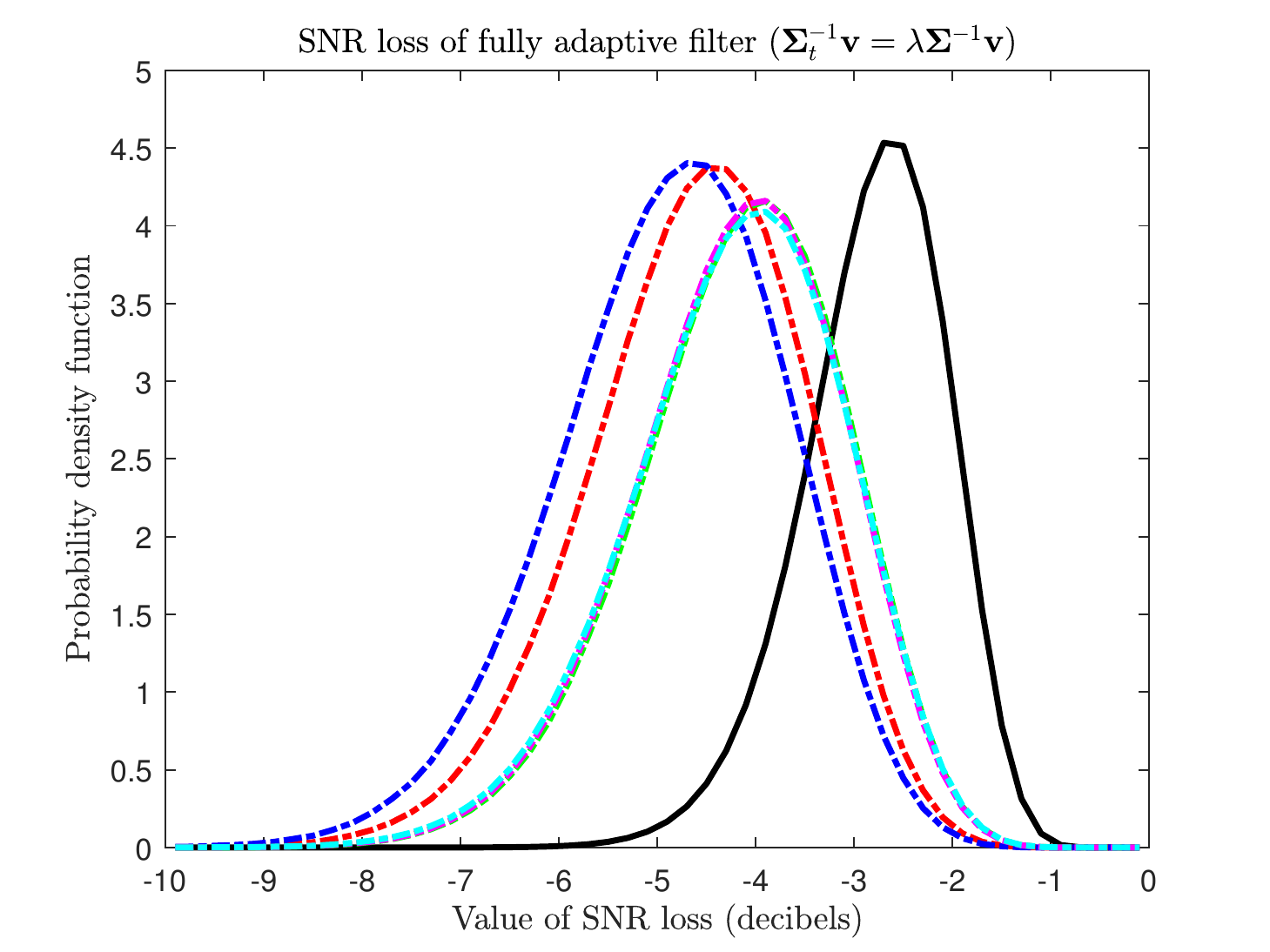}
	\caption{Distribution of the SNR loss when $\Ct \neq \C$ and  $\Ct^{-1}\vv=\lambda\C^{-1}\vv$. The solid line represents the case $\Ct=\C$. $N=16$ and $K=2N$.}
	\label{fig:pdf_snrloss_fa_GER_delta_mean_iW=6_K=32}
\end{figure}

\begin{figure}[h]
	\centering
	\includegraphics[width=11cm]{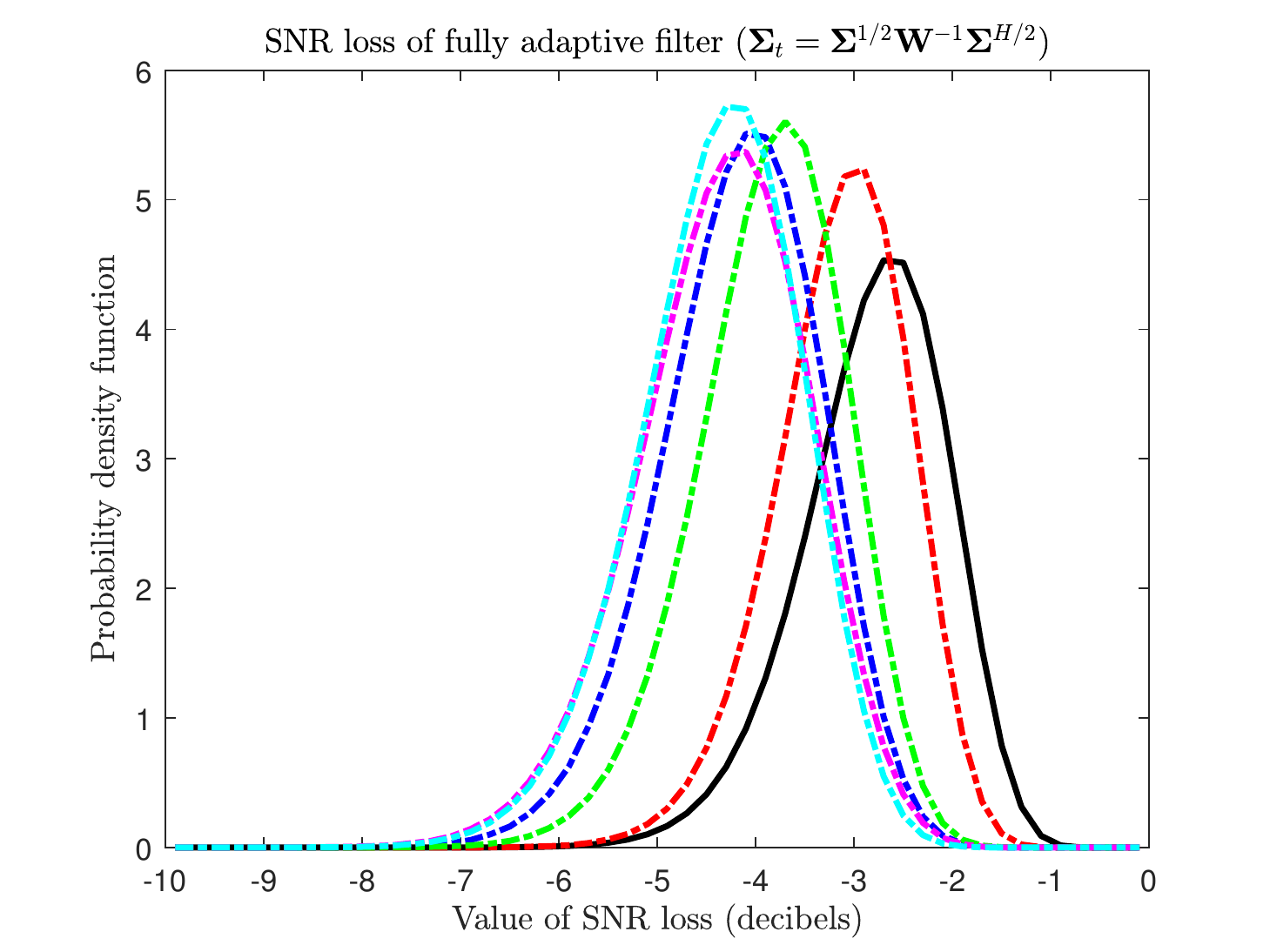}
	\caption{Distribution of the SNR loss when $\Ct=\C^{1/2}\W^{-1}\C^{H/2}$. The solid line represents the case $\Ct=\C$. $N=16$ and $K=2N$.}
	\label{fig:pdf_snrloss_fa_delta_mean_iW=6_K=32}
\end{figure}

Let us come back to the general case where $\Ct \neq \C$ but the GER $\Ct^{-1}\vv = \lambda \C^{-1}\vv$ is satisfied. The representation of the SNR loss is given by \eqref{rep_SNRloss_GER} and is shown to depend on the scenario through $\lambda$ and the eigenvalues $\lambda_{i}$. In the sequel we consider a scenario dealing with array processing, more precisely a uniform linear array with  $N=16$ elements spaced a half wavelength apart. The data to be filtered consists of thermal (white Gaussian) noise and $3$ interfering signals located at $-12^{\circ}$, $9^{\circ}$ and $25^{\circ}$ (measured with respect to the normal of the array), with respective powers $35$dB, $25$dB and $30$dB above thermal noise power.  We would like to stress that this is a very specific scenario where $\C$ consists of a very strong low-rank component plus a scaled identity matrix. Thus it may not be representative of other applications and consequently the conclusions drawn hereafter apply only to this case. As for $\Ct$ we will use the method in \cite{Besson20g} where  $\Ct=\C^{1/2}\W^{-1}\C^{H/2}$ and, in the general case, $\W$ is a Wishart matrix with mean value $\eta \eye{N}$ and $10\log_{10}\eta$ is uniformly distributed over [$-6$dB, $6$dB]. In order to enforce the GER the matrix $\W$ has a specific form, see \cite{Besson20g} for details.  In Figure \ref{fig:pdf_snrloss_fa_GER_delta_mean_iW=6_K=32} five independent matrices $\Ct$ were so randomly generated and we compare the distribution of the SNR loss when $\Ct \neq \C$ and  $\Ct^{-1}\vv=\lambda\C^{-1}\vv$ to the case where $\Ct=\C$ (solid line in the figure). As can be observed a mismatch of the training samples covariance matrix results in a degradation of the SNR loss.
Let us finally deal with the more general case where  $\Ct=\C^{1/2}\W^{-1}\C^{H/2}$ and the GER is not satisfied. Figure \ref{fig:pdf_snrloss_fa_delta_mean_iW=6_K=32} plots the distribution of the SNR loss corresponding to the representation in \eqref{rep_SNRloss_general}. Again the deleterious effect of $\Ct \neq \C$ is observed.

Before closing this section, we mention that approximations of the distribution of $\loss$ of the form
\begin{equation}\label{approx_representation_SNRloss}
\loss \approxdist \left[1 + a \frac{\Cchisquare{\nu}{0}}{\Cchisquare{\mu}{0}}\right]^{-1}
\end{equation}
are proposed in \cite{Besson20g} both under the GER assumption and in the general case. They are based on approximating a quadratic form in centered normal variables (GER case) or non centered Student variables (general case) and rely on on Pearson's method \cite{Pearson59}, see also \cite{Imhof61,Solomon77,Mathai92}. 

\subsection{The case of Student distributed training samples}
So far we assumed a fixed and arbitrary matrix $\Ct$ and, with no loss of generality, we can write it as $\Ct = \C^{1/2} \mGamma^{-1} \C^{H/2}$ where $\mGamma$ is an arbitrary matrix. Therefore, the p.d.f. of $\Xt | \mGamma$ is given by
\begin{align}\label{pdf_Xt|Gamma}
p(\Xt | \mGamma) &= \pi^{-NK} \det{\C^{1/2} \mGamma^{-1} \C^{H/2}}^{-K} \etr{-\Xt^{H}[\C^{1/2} \mGamma^{-1} \C^{H/2}]^{-1}\Xt} \nonumber \\
&= \pi^{-NK}  \det{\C}^{-K} \det{\mGamma}^{K} \etr{-\mGamma \C^{-1/2}\Xt\Xt^{H}\C^{-H/2}}
\end{align}
Let us now consider $\mGamma$ as a \emph{random matrix} and for mathematical convenience let us assume for $\mGamma$ a prior conjugate to \eqref{pdf_Xt|Gamma}, namely a Wishart distribution $\mGamma \dist \CW{N}{\nu}{\mu^{-1}\eye{N}}$
\begin{equation}\label{pdf_Gamma}
p(\mGamma)	\propto \det{\mu^{-1}\eye{N}}^{-\nu}\det{\mGamma}^{\nu-N} \etr{-\mu \mGamma}
\end{equation}
The mean value of $\mGamma$ is $\E{\mGamma}=(\nu-N)^{-1}\mu \eye{N}$ so that choosing $\mu=\nu-N$ results in $\mGamma$ ``fluctuating'' around $\eye{N}$ and hence $\Ct$ fluctuating around $\C$. Under the assumptions \eqref{pdf_Xt|Gamma}-\eqref{pdf_Gamma} it can be shown that $\Xt$ follows a complex matrix-variate Student distribution $\Xt \dist \mCT{N,K}{\nu-N+1}{\Mzero}{\mu\C}{\I_{K}}$ whose p.d.f is given by
\begin{equation}\label{pdf_Xt_Student}
p(\Xt) \propto \det{\mu\C}^{-K} \det{\I_{N}+(\mu\C)^{-1}\Xt\Xt^{H}}^{-(\nu+K)}	
\end{equation}
In other words this type of random covariance mismatch results in a distribution mismatch as the training samples are no longer Gaussian but Student distributed. In fact we have now $\Xt \dist (\mu\mSigma)^{\onehalf} \W_{\nu}^{-\onehalf}\N$ where $\W_{\nu} \dist \CW{N}{\nu}{\I_{N}}$ is independent of $\N \dist \mCN{N,K}{\mat{0}}{\I_{N}}{\I_{K}}$ so that
\begin{align}
\St 	\dist \mu \C^{\onehalf} \W_{\nu}^{-\onehalf} \W_{K} \W_{\nu}^{-\onehalf} \C^{\onehalf} =  \mu \C^{\onehalf} \F^{-1} \C^{\onehalf}
\end{align}
where $\W_{K} \dist \CW{N}{K}{\I_{N}}$. It follows that $\F=\W_{\nu}^{\onehalf} \W_{K}^{-1} \W_{\nu}^{\onehalf} \dist \CF{N}{\nu}{K}$. Recall that in the Gaussian case we had $\St \dist \Ct^{\onehalf} \W_{K} \Ct^{\onehalf}$ with $\W_{K} \dist \CW{N}{K}{\eye{N}}$. Therefore the analysis of the Student case follows along the same lines as in the Gaussian case except that now one needs to use properties of partitioned $F$-distributed matrices rather than partitioned Wishart matrices. In the Student case it can be shown \cite{BessonsnrlossT} that the SNR loss admits the following representation:
\begin{equation}\label{rep_SNRloss_Student}
\boxed{\lossStudent \dist \left[1 + \left(1+\frac{\Cchisquare{K-N+1}{0}}{\Cchisquare{\nu}{0}}\right) \frac{\Cchisquare{N-1}{0}}{\Cchisquare{K-N+2}{0}}\right]^{-1}}
\end{equation}
When comparing the Student representation \eqref{rep_SNRloss_Student} to its Gaussian counterpart \eqref{rep_SNRloss_MVDR} it is clear that the SNR loss is likely to take lower values in the Student case than in the Gaussian case. Note also that we recover that \eqref{rep_SNRloss_MVDR} and \eqref{rep_SNRloss_Student} are equivalent when $\nu \rightarrow \infty$. The p.d.f. of $\lossStudent$ is given by
\begin{equation}\label{pdf_SNRloss_Student}
\pdflossStudent(\loss) = 	\frac{\loss^{K-N+1} (1-\loss)^{N-2}}{\Beta{N-1}{K-N+2}} \times  \frac{\Beta{K-N+1}{\nu+N-1}}{\Beta{K-N+1}{\nu}} {\;}_{2}F_{1}(K+1,K-N+1;\nu+K;1-\loss)
\end{equation}
where the first term is recognized as the distribution of the SNR loss in the Gaussian case. 

\begin{figure}[h]
\centering
\includegraphics[width=11cm]{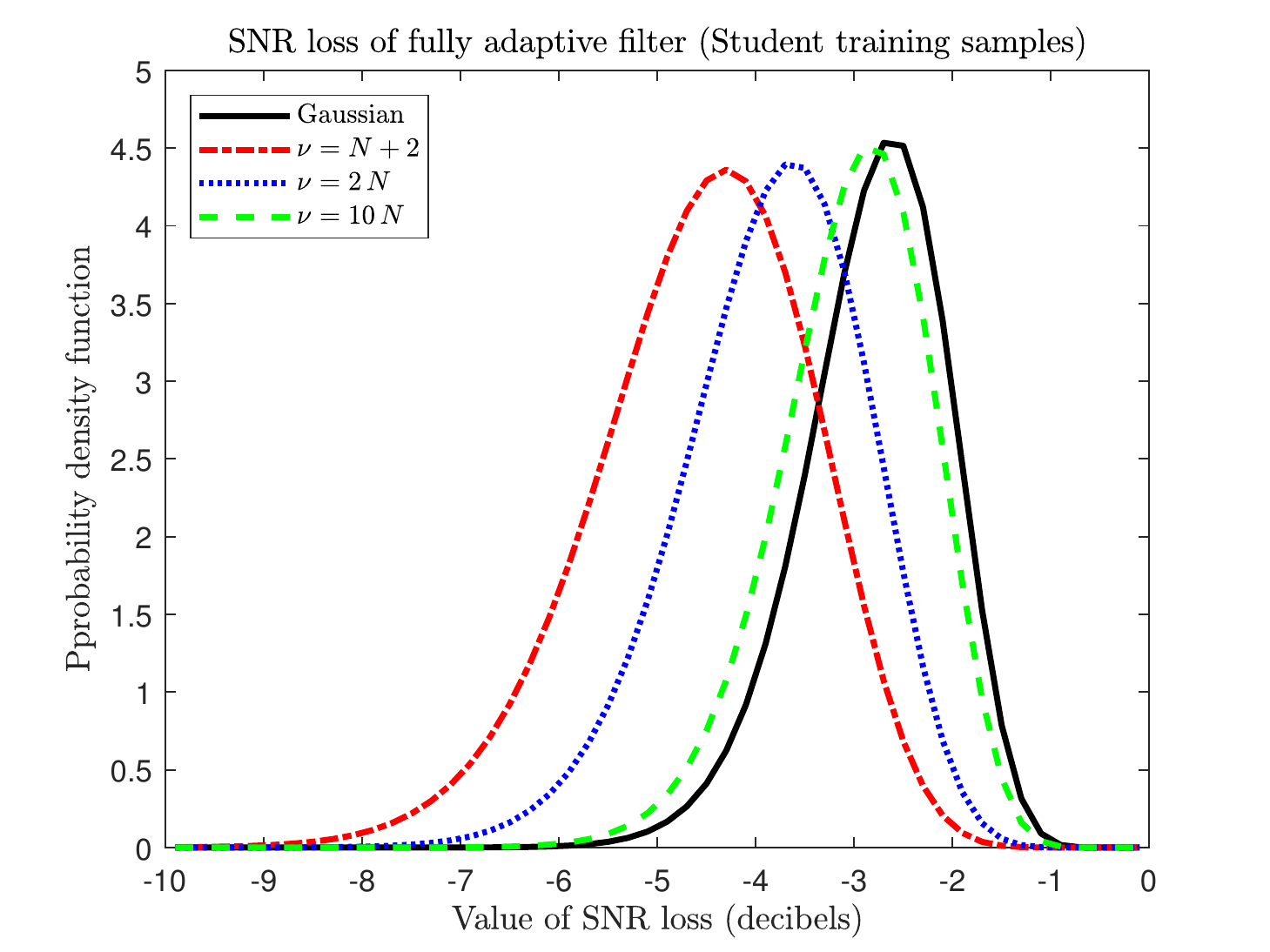} 
\caption{Probability density function of the SNR loss with Student distributed samples for various $\nu$. $\E{\Xt\Xt^{H}}=\C$. $N=16$ and $K=2N$.}
\label{fig:pdf_snrloss_fa_Student_K=32_mean_iW=1}
\end{figure}

\begin{figure}[h]
	\centering
	\includegraphics[width=11cm]{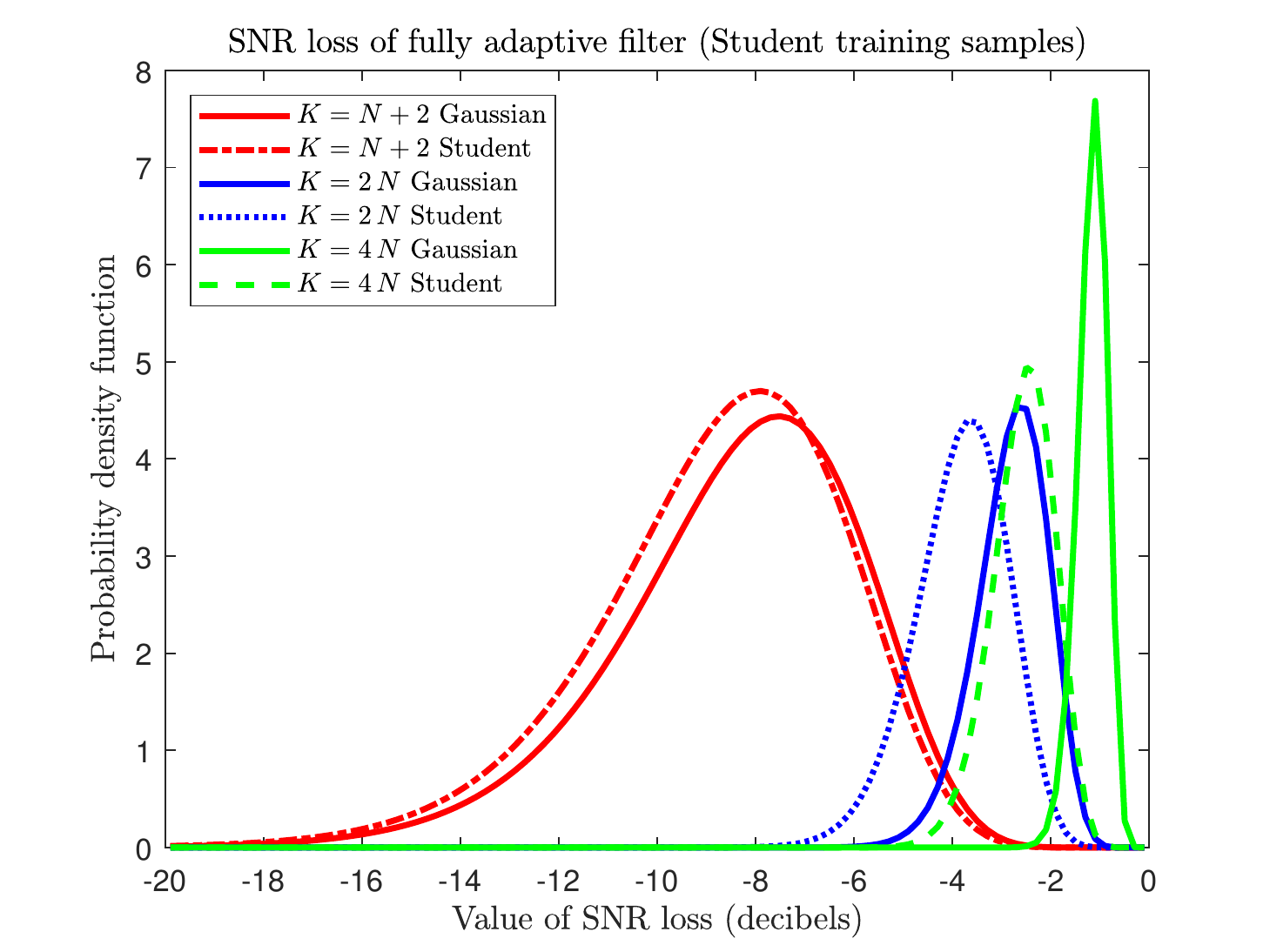} 
	\caption{Probability density function of the SNR loss with Student distributed samples for various $K$. $\E{\Xt\Xt^{H}}=\C$. $N=16$ and $\nu=2N$.}
	\label{fig:pdf_snrloss_fa_Student_nu=32_mean_iW=1}
\end{figure}

As an illustration of the impact of Student distributed training samples we  first look at the influence of $\nu$ in Figure \ref{fig:pdf_snrloss_fa_Student_K=32_mean_iW=1} where we display the p.d.f  of $\lossStudent$  for $K=2N$. Here $\mu=\nu-N$ so that $\E{\Xt\Xt^{H}}=\C$. As can be seen, the impact is rather significant with the support of $\lossStudent$'s p.d.f. moved downwards compared to that in the Gaussian case. We also observe that it depends however on $K$ as illustrated in Figure \ref{fig:pdf_snrloss_fa_Student_nu=32_mean_iW=1}: the difference between the Student and the Gaussian cases increases with $K$.

\subsection{Synthesis}
This section was devoted to the analysis of the SNR loss associated with the fully adaptive filter $\vw \propto \St^{-1}\vv$. We looked at the case where $\Ct = \C$ (for both Gaussian and Student distributed training samples) and also at the case $\Ct \neq \C$. A short synthesis of this section is found below.
\begin{synthesis}
\emph{When the training samples have covariance matrix $\Ct$ and the filter $\vw = (\vv^{H} \St^{-1} \vv)^{-1} \St^{-1} \vv$ is used in lieu of $\wopt =  (\vv^{H} \C^{-1} \vv)^{-1} \C^{-1} \vv$ degradation occurs with the SNR loss a random variable taking values in $[0,1]$. The least impact is achieved in the MVDR scenario where $\Ct=\C$ and $K=2N-3$ training samples are necessary to achieve an average SNR $3$dB below the optimum SNR. In the MPDR scenario where $\Ct=\C+P\vv\vv^{H}$ the number of samples required to achieve convergence increases with $\SNRopt$. The detrimental effect of a covariance mismatch between the training samples and the data to be filtered was also analysed. Finally we illustrated that with Student distributed training samples (which can occur as a particular (Bayesian) type of covariance mismatch) the support of the p.d.f. of $\loss$ is also moved towards lower values.}
\end{synthesis}

\subsection{Additional insights}
The SNR loss of the MPDR beamformer has been particularly studied in \cite{Boroson80} where errors on the signature $\vv$ are also considered. It allows a more straightforward derivation of $\lossmpdr$ which we adapt here. Let us assume that  $\Ct=\C+P\vv\vv^{H}$ and let us write  
\begin{align}\label{SNRloss_MPDR_direct}
\lossmpdr &= \frac{\left(\vv^{H} \St^{-1} \vv\right)^{2}}{\left(\vv^{H} \C^{-1} \vv\right) \left(\vv^{H} \St^{-1} \C\St^{-1} \vv\right)} \nonumber \\
&=\frac{\left(\vv^{H} \St^{-1} \vv\right)^{2}}{\left(\vv^{H} \Ct^{-1} \vv\right) \left(1+P\vv^{H} \C^{-1} \vv\right)}  \frac{1}{\left[\vv^{H} \St^{-1} \left( \Ct - P \vv \vv^{H}\right) \St^{-1} \vv\right]} \nonumber \\
&= \frac{\left(\vv^{H} \St^{-1} \vv\right)^{2}}{\left(\vv^{H} \Ct^{-1} \vv\right) \left(\vv^{H} \St^{-1} \Ct\St^{-1} \vv\right)} \left(1+P\vv^{H} \C^{-1} \vv\right)^{-1} \left(1- \frac{P \left(\vv^{H} \St^{-1} \vv\right)^{2}}{\vv^{H} \St^{-1} \Ct\St^{-1} \vv}\right)^{-1} \nonumber \\
&\dist \lossmvdr \left[ \left(1+P\vv^{H} \C^{-1} \vv\right) \left(1-\lossmvdr P \vv^{H} \Ct^{-1} \vv\right)\right]^{-1} \nonumber \\
&= \lossmvdr \left[ (1+P\vv^{H} \C^{-1} \vv - \lossmvdr P \vv^{H} \C^{-1} \vv\right]^{-1} \nonumber \\
&= \lossmvdr \left[ 1 + (1-\lossmvdr) \SNRopt \right]^{-1}
\end{align}
where we used the fact that $\vv^{H} \Ct^{-1} \vv = \left(1+P\vv^{H} \C^{-1} \vv\right)^{-1}  \left(\vv^{H} \C^{-1}\vv\right)$. The  simple change of variables $\lossmpdr \rightarrow \lossmvdr$ in \eqref{SNRloss_MPDR_direct}  along with the p.d.f. of $\lossmvdr$ in \eqref{pdf_SNRloss_MVDR} enables one to recover the p.d.f. of $\lossmpdr$ of \eqref{pdf_SNRloss_MPDR}.

The derivations above also enable one to derive \emph{stochastic representations for the weight vector} of the adaptive filter itself, see \cite{Richmond96} for related work. Indeed one has
\begin{align}
\vw &= \frac{\St^{-1} \vv}{\vv^{H} \St^{-1} \vv} \nonumber \\
&\dist \frac{\Ct^{-H/2} \W^{-1} \Ct^{-1/2} \vv}{\vv^{H} \Ct^{-H/2} \W^{-1} \Ct^{-1/2} \vv} \nonumber \\
&\dist \frac{\Ct^{-H/2} \Q \W^{-1} \Q^{H} \Ct^{-1/2} \vv}{\vv^{H} \Ct^{-H/2} \Q \W^{-1} \Q^{H}\Ct^{-1/2} \vv} \nonumber \\
\end{align}
for any unitary $\Q$. Let us again choose $\Q = \begin{bmatrix} \frac{\Ct^{-1/2}\vv}{(\vv^{H}\Ct^{-1}\vv)^{1/2}} & \Ct^{H/2}\Vorth (\Vorth^{H}\Ct\Vorth)^{-H/2} \end{bmatrix}$ where $\Vorth$ is a semi-unitary matrix such that $\Vorth^{H}\vv=\vzero$. Then
\begin{align}
\vw & \dist (\vv^{H}\Ct^{-1}\vv)^{-1/2} \frac{\Ct^{-H/2}\Q\W^{-1}\efirst}{\efirst^{H}\W^{-1}\efirst} \nonumber \\
&= (\vv^{H}\Ct^{-1}\vv)^{-1/2} \begin{bmatrix} \frac{\Ct^{-1}\vv}{(\vv^{H}\Ct^{-1}\vv)^{1/2}} & \Vorth (\Vorth^{H}\Ct\Vorth)^{-H/2} \end{bmatrix} \begin{bmatrix} 1 \\ - \vt_{21} \end{bmatrix} \nonumber \\
&= \frac{\Ct^{-1} \vv}{\vv^{H} \Ct^{-1} \vv} - (\vv^{H}\Ct^{-1/2}\vv)^{-1/2} \Vorth (\Vorth^{H}\Ct\Vorth)^{-H/2} \vt_{21} \nonumber \\
&\dist \frac{\Ct^{-1} \vv}{\vv^{H} \Ct^{-1} \vv} - (\vv^{H}\Ct^{-1}\vv)^{-1/2} \Vorth (\Vorth^{H}\Ct\Vorth)^{-H/2} \frac{\vn_{21}}{\sqrt{V_{21}}}
\end{align}
with $\vn_{21} \dist \vCN{N-1}{\vzero}{\eye{N-1}}$ and $V_{21} \dist \Cchisquare{K-N+2}{0}$. When $\Ct=\C$ we obtain
\begin{equation}\label{rep_wmvdr} 
\wmvdr \dist \wopt - \frac{1}{ (\vv^{H}\C^{-1}\vv)^{1/2}}  \Vorth (\Vorth^{H}\C\Vorth)^{-H/2} \frac{\vn_{21}}{\sqrt{V_{21}}}
\end{equation}
while when $\Ct=\C+P\vv\vv^{H}$ we get
\begin{equation}\label{rep_wpvdr} 
\wmpdr \dist \wopt - \frac{(1+\SNRopt)^{1/2}}{ (\vv^{H}\C^{-1}\vv)^{1/2}}  \Vorth (\Vorth^{H}\C\Vorth)^{-H/2}\frac{\vn_{21}}{\sqrt{V_{21}}}
\end{equation}
We can notice that the two beamformers differ in the subspace orthogonal to $\vv$. One can also observe that
\begin{equation}\label{average_snorm_w}
\E{\sqnorm{\vw}} = \sqnorm{\frac{\Ct^{-1} \vv}{\vv^{H} \Ct^{-1} \vv}} +  \frac{1}{ \vv^{H}\Ct^{-1}\vv} \frac{\Tr{(\Vorth^{H} \Ct \Vorth)^{-1}}}{K-N+1}
\end{equation}
This implies in particular that
\begin{equation}\label{wnag_mvdr_mpdr}
\E{\sqnorm{\wmpdr}} = \E{\sqnorm{\wmvdr}} + P \frac{\Tr{\left(\Vorth^{H} \C \Vorth\right)^{-1}}}{K-N+1}
\end{equation}
Therefore we can expect the white noise array gain $\isqnorm{\wmpdr}$ of the MPDR beamformer to be lower than that of the MVDR beamformer, with a difference that increases with the power of the signal of interest.

\section{Partially adaptive array processing}
This section is devoted to partially adaptive processing where the adaptive filter belongs to a subspace of the entire space. In array processing applications this includes beamspace processing or reduced-rank adaptive beamforming. First we set the principle of this approach and give some insights about when it can be as efficient as fully adaptive processing. Then we will successively consider the case of fixed transformations and the case of principal-component based transformations.  We will also briefly allude to random transformations. Unless otherwise stated we assume in this section that $\Ct=\C$.
\subsection{Structure of and motivation for partially adaptive array processing}
\begin{figure}[h]
	\centering
	\includegraphics[width=8cm]{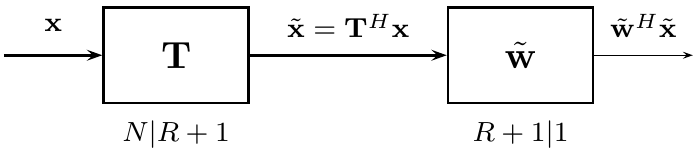} 
	\caption{Structure of a partially adaptive processor with $\T$ a $N|(R+1)$ matrix.}
	\label{fig:structure_pa_df}
\end{figure}

Consider the general structure of a partially adaptive beamformer of the form described in Figure \ref{fig:structure_pa_df}. 

\begin{figure}[h]
	\centering
	\includegraphics[width=10cm]{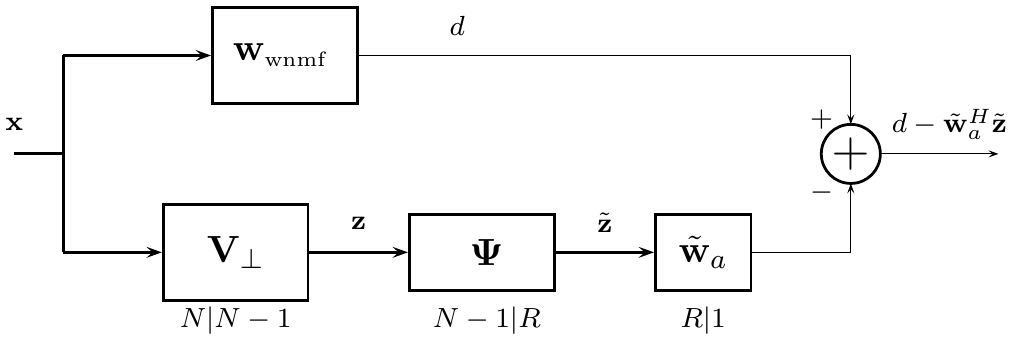} 
	\caption{Structure of a partially adaptive processor with $\T = \begin{bmatrix} \wcbf & \Vorth \mPsi \end{bmatrix}$.}
	\label{fig:structure_pa_gsc}
\end{figure}

The data $\vx$ is first transformed through $\T$ to get $\vxtilde = \T^{H}\vx$ and then $\vxtilde$ is filtered by the length-$(R+1)$ filter $\vwtilde$ to obtain the output. The equivalent  length-$N$ filter is $\vw = \T \vwtilde$.  We will often consider the case where $\T = \begin{bmatrix} \isqnorm{\vv} \vv & \Vorth \mPsi \end{bmatrix}$ with $\mPsi$ a $(N-1)|R$ matrix, which is described in Figure \ref{fig:structure_pa_gsc}.

This choice of $\T$ is reminiscent of a sidelobe canceler structure where the columns of $\Vorth \mPsi$ are orthogonal to $\vv$ and aimed at capturing the part of noise that goes through the white noise matched filter $\wcbf = \isqnorm{\vv} \vv$.

Let us start as before with minimizing the output power subject to unit gain towards $\vv$:
\begin{equation}\label{pb_pa_mvdr}
\underset{\vwtilde}{\min} \; \vwtilde^{H} \Cttilde \vwtilde \text{ subject to } \vwtilde^{H}\vvtilde=1	
\end{equation}
where $\Cttilde = \T^{H}\Ct\T = \T^{H}\C\T$ and $\vvtilde=\T^{H}\vv$. The solution is $\wtildeopt = (\vvtilde^{H}\Cttilde^{-1}\vvtilde)^{-1}\Cttilde^{-1}\vvtilde$. 

Let us now ask the following question: 
\begin{center}
\begin{tcolorbox}[colback=white,colframe=green!20!blue,width=0.7\textwidth]
Can we possibly have $\T\wtildeopt = \wopt  = (\vv^{H}\C^{-1}\vv)^{-1}\C^{-1}\vv$? 
\end{tcolorbox}
\end{center}
If so it would mean that no loss is incurred when using a partially adaptive processor compared to a fully adaptive processor. Let us examine how such an equality could be achieved:
\begin{align}\label{condition_PA=FA}
\T\wtildeopt\propto \C^{-1}\vv & \Leftrightarrow \T\Cttilde^{-1}\vvtilde \propto \C^{-1}\vv \nonumber \\
&\Leftrightarrow \T(\T^{H}\C\T)^{-1} \T^{H}\vv \propto \C^{-1}\vv \nonumber \\
&\Leftrightarrow \C^{\onehalf}\T(\T^{H}\C\T)^{-1}\T^{H}\C^{\onehalf} \C^{-\onehalf}\vv \propto \C^{-\onehalf}\vv \nonumber \\
&\Leftrightarrow \Proj{\C^{\onehalf}\T} \C^{-\onehalf}\vv \propto \C^{-\onehalf}\vv \nonumber \\
&\Leftrightarrow \C^{-\onehalf}\vv  \in \range{\C^{\onehalf}\T}\nonumber \\
&\Leftrightarrow \C^{-1}\vv  \in \range{\T}\nonumber \\
\end{align}
The latter equality means that the range space of $\T$ should include $\C^{-1}\vv$. At first glance this is a meaningless condition since, if $\C^{-1}\vv$ was known, we would already get the optimal filter. However, let us consider a situation where $\C = \G\G^{H} + \gamma\eye{N}$ where $\G$ is a $N|R$ matrix. Then
\begin{align}\label{wopt_lowrank+white}
\C^{-1}\vv &= (\G\G^{H} + \gamma\eye{N})^{-1}\vv \nonumber \\
&= \gamma^{-1}\left[\eye{N} - \G \left(\gamma\eye{R}+\G^{H}\G \right)^{-1} \G^{H} \right]  \vv \nonumber \\
&= \gamma^{-1} \begin{bmatrix}\vv & \G \end{bmatrix} \begin{bmatrix} 1 \\ -\left(\gamma\eye{R}+\G^{H}\G \right)^{-1}\G^{H}\vv \end{bmatrix}
\end{align}
It follows that if $\begin{bmatrix}\vv & \G \end{bmatrix} \in \range{\T} $ then $\T\wtildeopt = \wopt$. In other words, \emph{if the range space $\T$ contains $\vv$ and the principal subspace of $\C$ then there is no loss in using partially adaptive processing}. In this case $\wopt$ actually belongs to a subspace. This suggests that partially adaptive processing might be particularly effective when the noise covariance matrix has a strong low-rank component. Before pursuing let us just see what the condition \eqref{condition_PA=FA} means when $\T = \begin{bmatrix} \isqnorm{\vv} \vv & \Vorth \mPsi \end{bmatrix}$. We have
\begin{align}\label{wopt_approx__lowrank+white}
\C^{-1}\vv  \in \range{\T} &\Leftrightarrow 	\C^{-1}\vv = \alpha \wcbf + \Vorth \mPsi \vbeta \nonumber \\
&\Leftrightarrow (\isqnorm{\vv} \vv \vv^{H} + \Vorth\Vorth^{H})\C^{-1}\vv  = \alpha \isqnorm{\vv} \vv + \Vorth \mPsi \vect{\beta} \nonumber \\
&\Leftrightarrow \isqnorm{\vv} (\vv^{H}\C^{-1}\vv-\alpha) \vv  + \Vorth (\Vorth^{H}\C^{-1}\vv - \mPsi \vbeta ) = \vzero 
\end{align}
Since $\vv$ and $\Vorth$ are orthogonal this is equivalent to $\Vorth^{H}\C^{-1}\vv \in \range{\mPsi}$. 

Moreover, if $\C = \G\G^{H} + \gamma\eye{N}$ then from \eqref{wopt_lowrank+white} $\Vorth^{H}\C^{-1}\vv = -\gamma^{-1} \Vorth^{H}\G \left(\gamma\eye{R}+\G^{H}\G \right)^{-1}  \G^{H}\vv \in \range{\Vorth^{H}\G}$. Therefore \emph{if $\range{\Vorth^{H}\G} \subset \range{\mPsi}$ the partially adaptive filter coincides with the fully adaptive filter}.

Before continuing note that if $\C = \G\G^{H} + \gamma\eye{N}$ and $\G\G^{H} \gg \gamma\eye{N}$ then from \eqref{wopt_lowrank+white}
\begin{equation}
\wopt \propto  \vv  - \G \left(\gamma\eye{R}+\G^{H}\G \right)^{-1} \G^{H}   \vv \simeq \vv  - \G \left(\G^{H}\G \right)^{-1} \G^{H}   \vv = \Projorth{\G}\vv
\end{equation}
where $\Projorth{\G}$ denotes the projector onto the orthogonal complement of $\range{\G}$. Therefore if the low-rank component in $\C$ is much stronger than the white noise component, the optimal filter consists of projecting the desired signal $\vv$ onto the so-called noise subspace. This property will be the basis for principal component adaptive processing to be described later.

\subsection{Analysis of the SNR loss for fixed $\T$ and insights}
Let us now consider the practical case where a set of training samples $\Xt$ is used to design the filter $\vwtilde$ and we solve \eqref{pb_pa_mvdr} with  $\Cttilde$ substituted for $\Sttilde = \T^{H} \St \T$, i.e.,
\begin{equation}
\underset{\vwtilde}{\min} \; \vwtilde^{H} \Sttilde \vwtilde \text{ subject to } \vwtilde^{H}\vvtilde=1	
\end{equation}
whose solution is
\begin{equation}
\vwtilde = \frac{\Sttilde^{-1}\vvtilde}{\vvtilde^{H}\Sttilde^{-1}\vvtilde}	 = \frac{(\T^{H} \St \T)^{-1}\T^{H}\vv}{\vv^{H}\T(\T^{H} \St \T)^{-1}\T^{H}\vv} \Rightarrow \vw = \frac{\T(\T^{H} \St \T)^{-1}\T^{H}\vv}{\vv^{H}\T(\T^{H} \St \T)^{-1}\T^{H}\vv}
\end{equation}
In the case where $\T=\begin{bmatrix} \wcbf & \Vorth \mPsi \end{bmatrix}$ the weight vector can be written as $\vw = \wcbf - \Vorth \mPsi \vwtildea$ with $\vwtildea$ a $R$-dimensional vector which is found by minimizing the output power in an unconstrained way since $(\wcbf - \Vorth \mPsi \vwtildea)^{H}\vv=1$. In other words the problem is now
\begin{equation}\label{pb_pa_mvdr_gsc}
\underset{\vwtildea}{\min} \; 	(\wcbf-\Vorth \mPsi\vwtildea)^{H} \St (\wcbf-\Vorth \mPsi\vwtildea)
\end{equation}
whose solution results in the filter
\begin{align}\label{sol_pa_mvdr_gsc}
\vw &= \wcbf - \Vorth \mPsi (\mPsi^{H}\Vorth^{H}\St\Vorth\mPsi)^{-1}\mPsi^{H}\Vorth^{H}\St\wcbf \nonumber \\
&= \begin{bmatrix} \wcbf & \Vorth \mPsi \end{bmatrix} \begin{bmatrix} 1 \\ -\Chat_{\vztilde}^{-1}\hat{r}_{d\vztilde}\end{bmatrix}
\end{align}
where $\Chat_{\vztilde}$ and $\hat{r}_{d\vztilde}$ are the sample versions of $\E{\vztilde \vztilde^{H}}$ and $\E{d^{\ast} \vztilde}$, see Figure \ref{fig:structure_pa_gsc}. 

We first analyse the SNR loss obtained with the filter $\vw = \T \vwtilde = (\vvtilde^{H}\Sttilde^{-1}\vvtilde)^{-1} \T\Sttilde^{-1}\vvtilde$. Let us assume first an arbitrary $\Ct$ and note that $\Xttilde = \T^{H} \Xt \dist \mCN{R+1,K}{\Mzero}{\Cttilde}{\eye{K}}$ so that $\Sttilde =\Xttilde\Xttilde^{H} \dist  \Cttilde^{1/2} \Wtilde \Cttilde^{H/2} \T$ with $\Wtilde \dist \CW{R+1}{K}{\eye{R+1}}$. Mimicking the derivations used in the fully adaptive case, we can write that
\begin{align}\label{deriv_rep_rho_pa}
\loss  &= \frac{(\vw^H \vv)^{2}}{(\vv^{H}\C^{-1}\vv)(\vw^{H}\C\vv)} \nonumber \\
&= \frac{(\vvtilde^H \Sttilde^{-1} \vvtilde)^{2}}{(\vv^{H}\C^{-1}\vv)(\vvtilde^{H}\Sttilde^{-1}\Ctilde\Sttilde^{-1}\vvtilde)} \nonumber \\
&\dist  \frac{(\vvtilde^H \Cttilde^{-H/2} \Wtilde^{-1} \Cttilde^{-1/2}  \vvtilde)^{2}}{(\vv^{H}\C^{-1}\vv)(\vvtilde^{H}\Cttilde^{-H/2}  \Wtilde^{-1}  \Cttilde^{-1/2} \Ctilde \Cttilde^{-H/2} \Wtilde^{-1}  \Cttilde^{-1/2} \vvtilde)} \nonumber \\
&\dist  \frac{(\vvtilde^H \Cttilde^{-H/2} \Qtilde \Wtilde^{-1} \Qtilde^{H} \Ct^{-1/2}  \vvtilde)^{2}}{(\vv^{H}\C^{-1}\vv)(\vvtilde^{H}\Cttilde^{-H/2} \Qtilde  \Wtilde^{-1}  \Qtilde^{H} \Cttilde^{-1/2} \Ctilde \Cttilde^{-H/2} \Qtilde \Wtilde^{-1} \Qtilde^{H}  \Cttilde^{-1/2} \vv)}
\end{align}
for any unitary matrix $\Qtilde$. Again let us choose $\Qtilde$ such that $\Qtilde^{H}  \Cttilde^{-1/2} \vvtilde = (\vvtilde^{H}\Cttilde^{-1}\vvtilde)^{1/2} \etildefirst$ where $\etildefirst = \begin{bmatrix} 1 & 0 & \ldots & 0 \end{bmatrix}^{T}$ and let us define
\begin{align}\label{Omega_tilde}
\mOmegatilde &= \Qtilde^{H} \Cttilde^{-1/2} \Ctilde \Cttilde^{-H/2} \Qtilde  = \begin{pmatrix} \tilde{\Omega}_{11} & \mOmegatilde_{12} \\ \mOmegatilde_{21} & \mOmegatilde_{22} \end{pmatrix}
\end{align}
Then  we get
\begin{equation} 
\loss 	\dist  \frac{\vvtilde^{H}\Cttilde^{-1}\vvtilde}{\vv^{H}\C^{-1}\vv} \times \frac{(\etildefirst^{H}\Wtilde^{-1}\etildefirst)^{2}}{\etildefirst^{H}\Wtilde^{-1}\mOmegatilde\Wtilde^{-1}\etildefirst}
\end{equation}
Similarly to what was done in the fully adaptive case, one can show that
\begin{equation}
\frac{\etildefirst^{H}\Wtilde^{-1}\mOmegatilde\Wtilde^{-1}\etildefirst}	{(\etildefirst^{H}\Wtilde^{-1}\etildefirst)^{2}} = \frac{\vvtilde^{H}\Cttilde^{-1}\vvtilde}{\vvtilde^{H}\Ctilde^{-1}\vvtilde} + \tilde{Q} = \frac{\vvtilde^{H}\Cttilde^{-1}\vvtilde}{\vvtilde^{H}\Ctilde^{-1}\vvtilde} \left[1+\frac{\vvtilde^{H}\Ctilde^{-1}\vvtilde}{\vvtilde^{H}\Cttilde^{-1}\vvtilde}\tilde{Q}\right]
\end{equation}
with
\begin{equation}
\tilde{Q}	\dist \tilde{V}_{21}^{-1} \sum_{i=1}^{R} \tilde{\lambda}_{i} \Cchisquare{1}{\tilde{V}_{21} \tilde{\delta}_{i}}
\end{equation}
where $\tilde{V}_{21} \dist \Cchisquare{K-R+1}{0}$, $\tilde{\delta}_{i}=| \tilde{\vu}_{i}^{H}\mOmegatilde_{22}^{-1}\mOmegatilde_{21}|^{2}$ and $\tilde{\vu}_{i}$, $\tilde{\lambda}_{i}$ are the $R$ eigenvectors and eigenvalues of $\mOmegatilde_{22}$. It follows that, for arbitrary $\Ct$, the SNR loss of the partially adaptive filter can be represented as
\begin{equation}\label{rep_SNRloss_PA_general}
\loss 	\dist  \frac{\vvtilde^{H}\Ctilde^{-1}\vvtilde}{\vv^{H}\C^{-1}\vv} 	\left[1+\frac{\vvtilde^{H}\Ctilde^{-1}\vvtilde}{\vvtilde^{H}\Cttilde^{-1}\vvtilde} \frac{\sum_{i=1}^{R} \tilde{\lambda}_{i} \Cchisquare{1}{\tilde{V}_{21} \tilde{\delta}_{i}}}{\tilde{V}_{21}}\right]^{-1}
\end{equation}
\emph{The previous equation provides the statistical representation of the SNR loss for any $\C$, $\Ct$ and $\T$}.

\begin{figure}[htb]
\centering
\subfigure[$N=16$, $K=16$, $R=4$]{
\includegraphics[width=7.5cm]{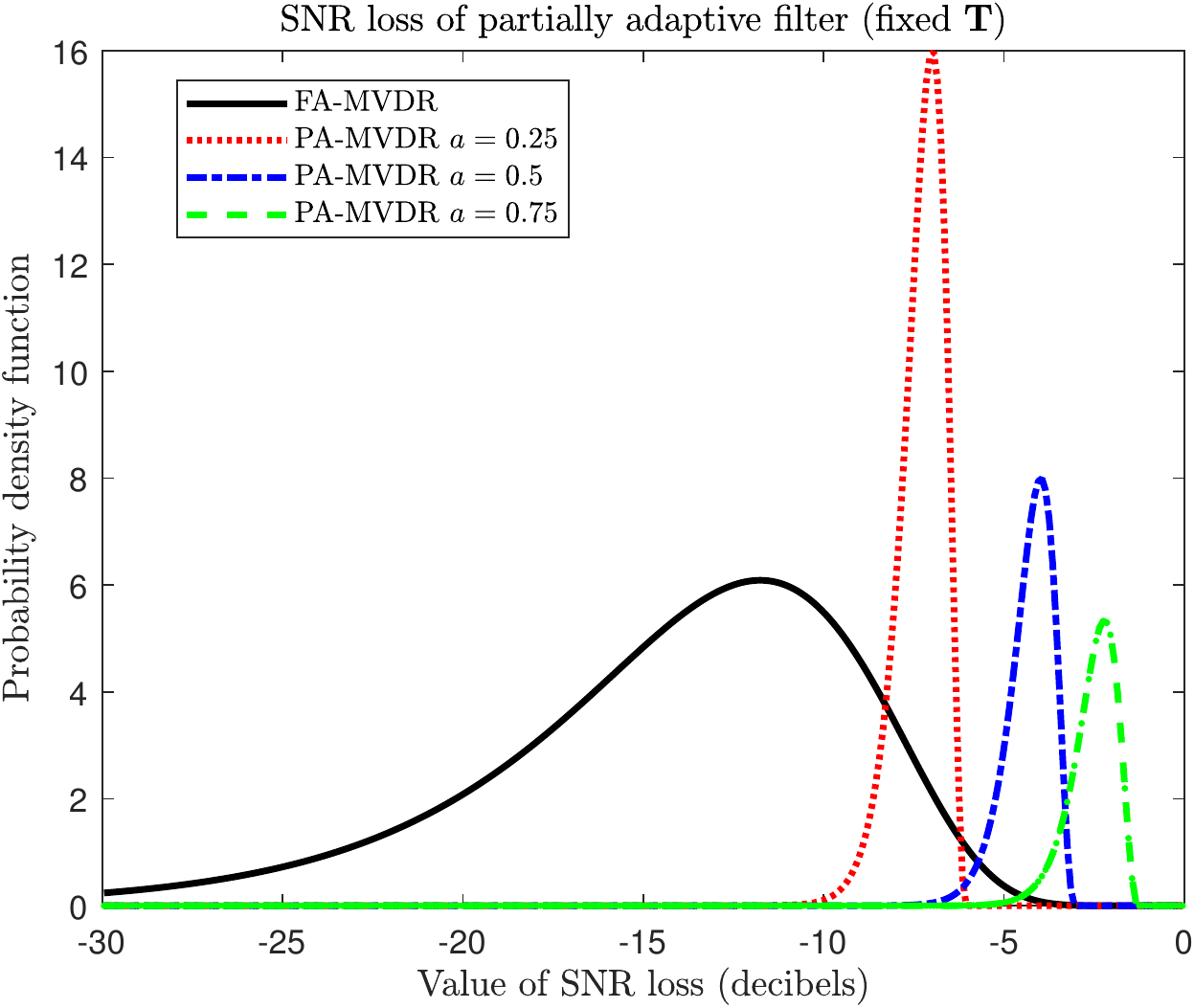}}
\subfigure[$N=16$, $K=32$, $R=4$]{%
\includegraphics[width=7.5cm]{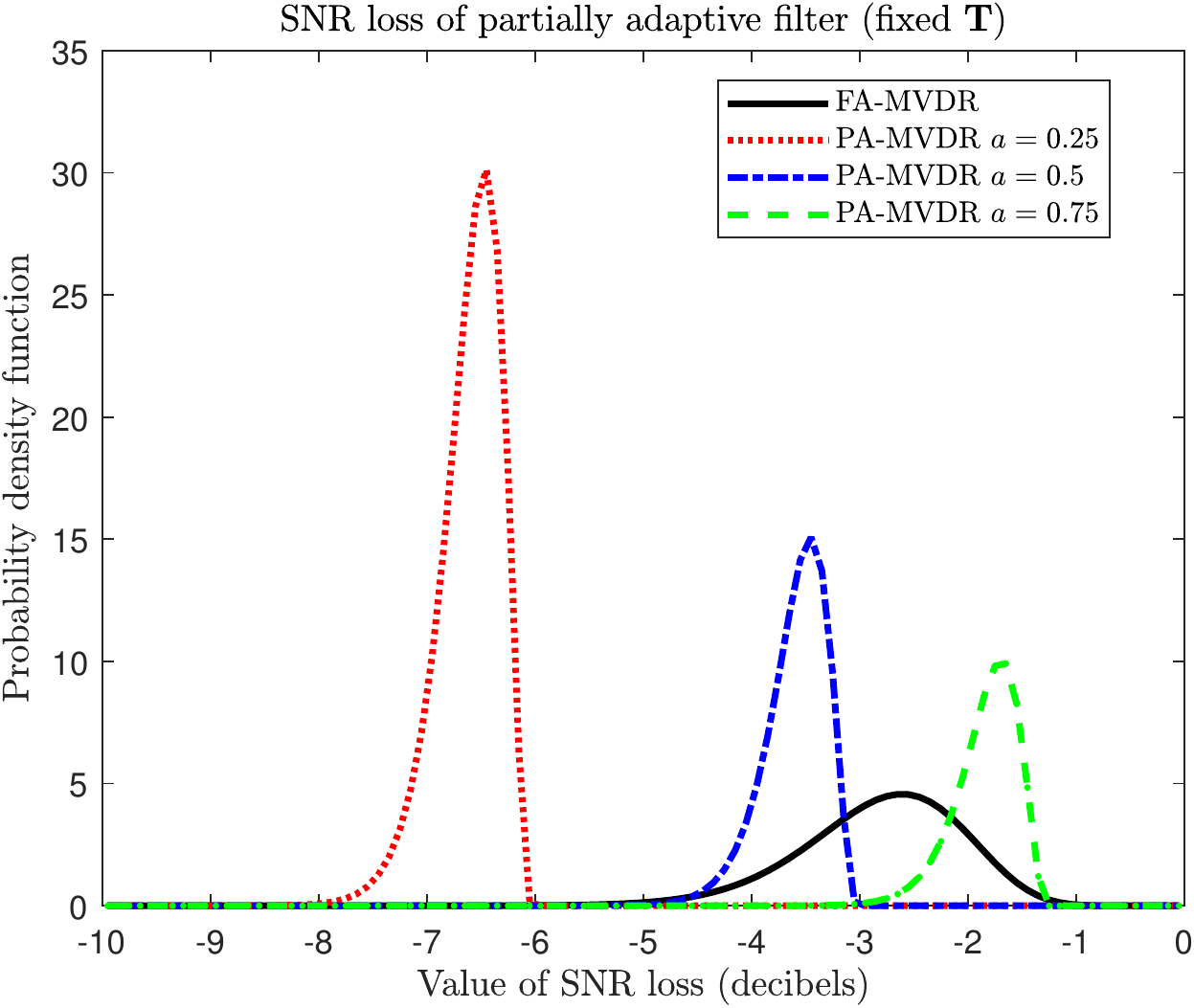}}\\
\subfigure[$N=64$, $K=64$, $R=16$]{%
\includegraphics[width=7.5cm]{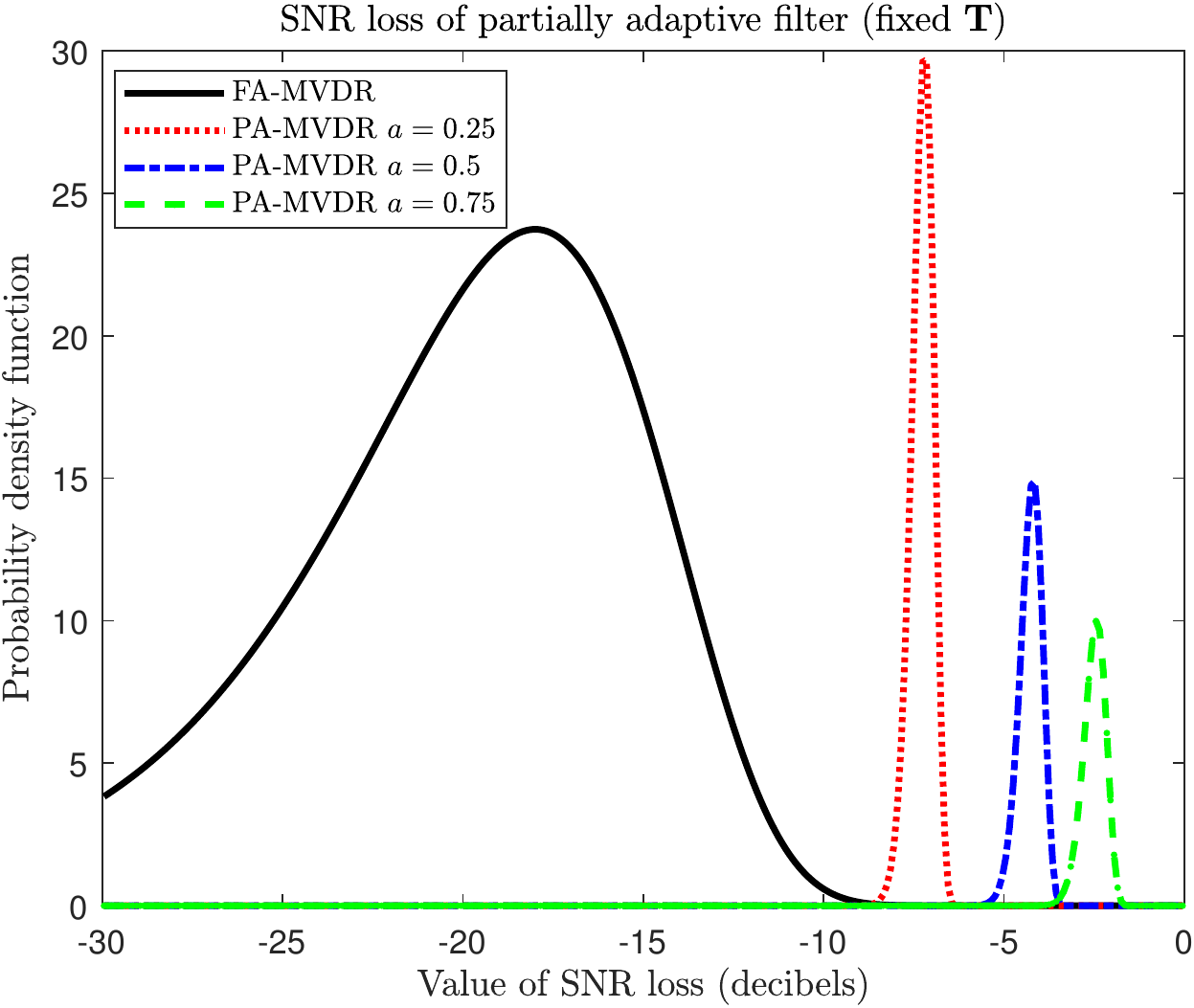}}
\subfigure[$N=64$, $K=128$, $R=164$]{%
\includegraphics[width=7.5cm]{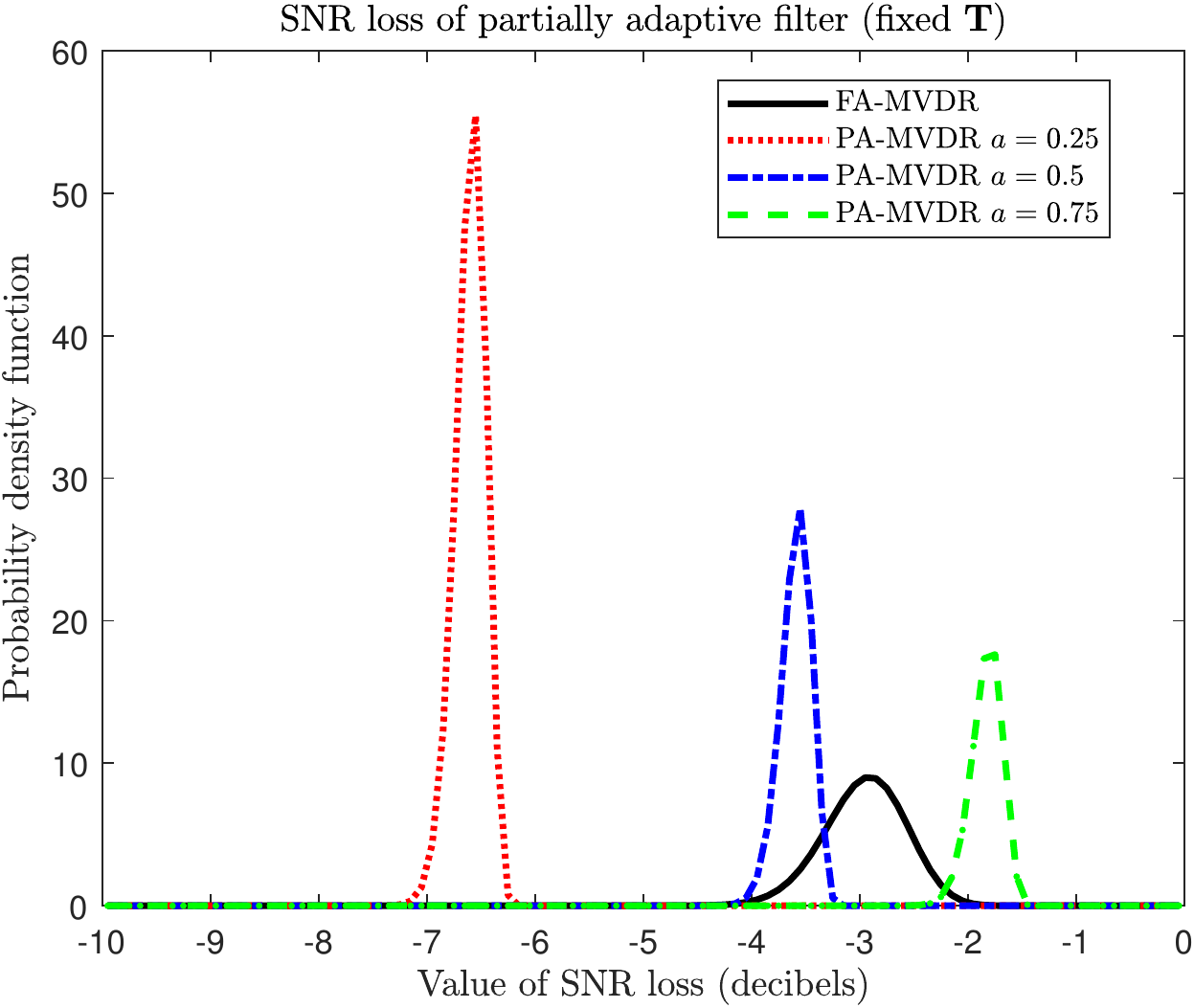}}
\caption{Distribution of the SNR loss of a partially adaptive filter versus value of $a$. The left panels deal with $K=N$, the right panels with $K=2N$. $R=N/4$.}
\label{fig:pdf_sinrloss_PA-MVDR}
\end{figure}

Let us now focus on the MVDR scenario for which $\Ct=\C$. In this case $\mOmegatilde=\eye{R+1}$, $\tilde{\lambda}_{i}=1$, $\tilde{\delta}_{i}=0$ and therefore one obtains
\begin{equation}\label{rep_SNRloss_PA-MVDR}
\boxed{\losspamvdr \dist 	\frac{\vvtilde^{H}\Ctilde^{-1}\vvtilde}{\vv^{H}\C^{-1}\vv}  \betapdf{R}{K-R+1}}
\end{equation}
Therefore, \emph{the SNR loss is now a scaled beta distributed random variable} whose p.d.f. greatly depends on
\begin{align}\label{coef_snrloss_pa}
a &= \frac{\vvtilde^{H}\Ctilde^{-1}\vvtilde}{\vv^{H}\C^{-1}\vv} = \frac{\vv^{H}\T(\T^{H}\C\T)^{-1}\T^{H}\vv}{\vv^{H}\C^{-1}\vv}	 = \frac{\vv^{H}\C^{-H/2} \Proj{\C^{1/2}\T}\C^{-1/2}\vv}{\vv^{H}\C^{-1}\vv} \nonumber \\
&= \frac{\text{energy of }\C^{-1/2}\vv \text{ in } \range{\C^{1/2}\T}}{\text{energy of }\C^{-1/2}\vv}
\end{align}
which confirms what was observed in \eqref{condition_PA=FA}, i.e., the importance of how much energy of $\C^{-1/2}\vv$ is contained in $\range{\C^{1/2}\T}$. In particular the support of the distribution of $\loss$ is shrunk due to the scaling and is now limited to $[0,a]$. Of course the value of $a$ is unknown since it depends on $\C$. In Figure \ref{fig:pdf_sinrloss_PA-MVDR} we illustrate the distribution of $\losspamvdr$ for various values of $N$, $K$ and an hypothetical $a$. It can be seen that in limited sample support, e.g., $K=N$ then partially adaptive beamforming is even to be preferred to fully adaptive beamforming. 

\begin{figure}[htb]
\centering
\subfigure{
\includegraphics[width=7.5cm]{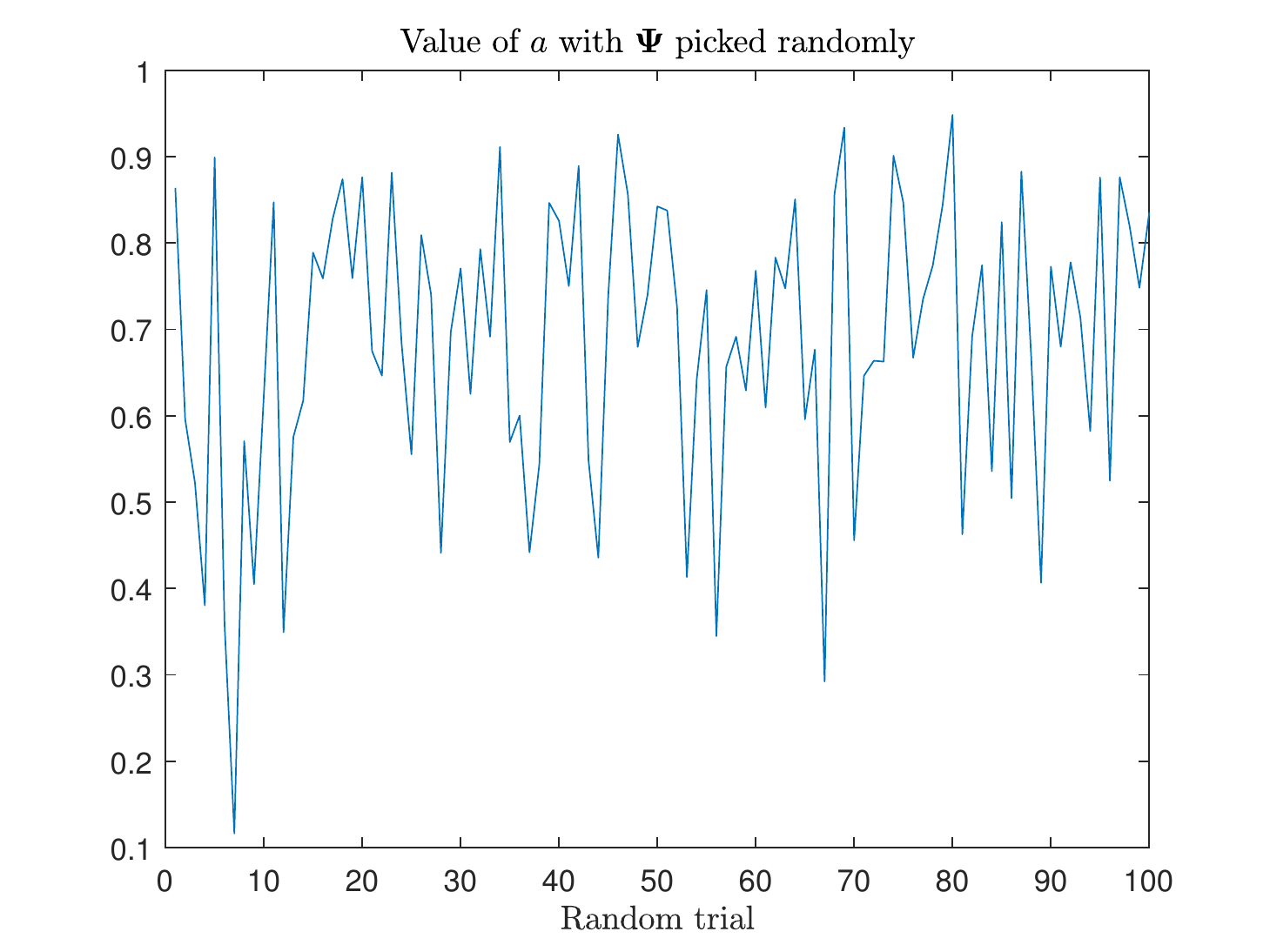}}
\subfigure{%
\includegraphics[width=7.5cm]{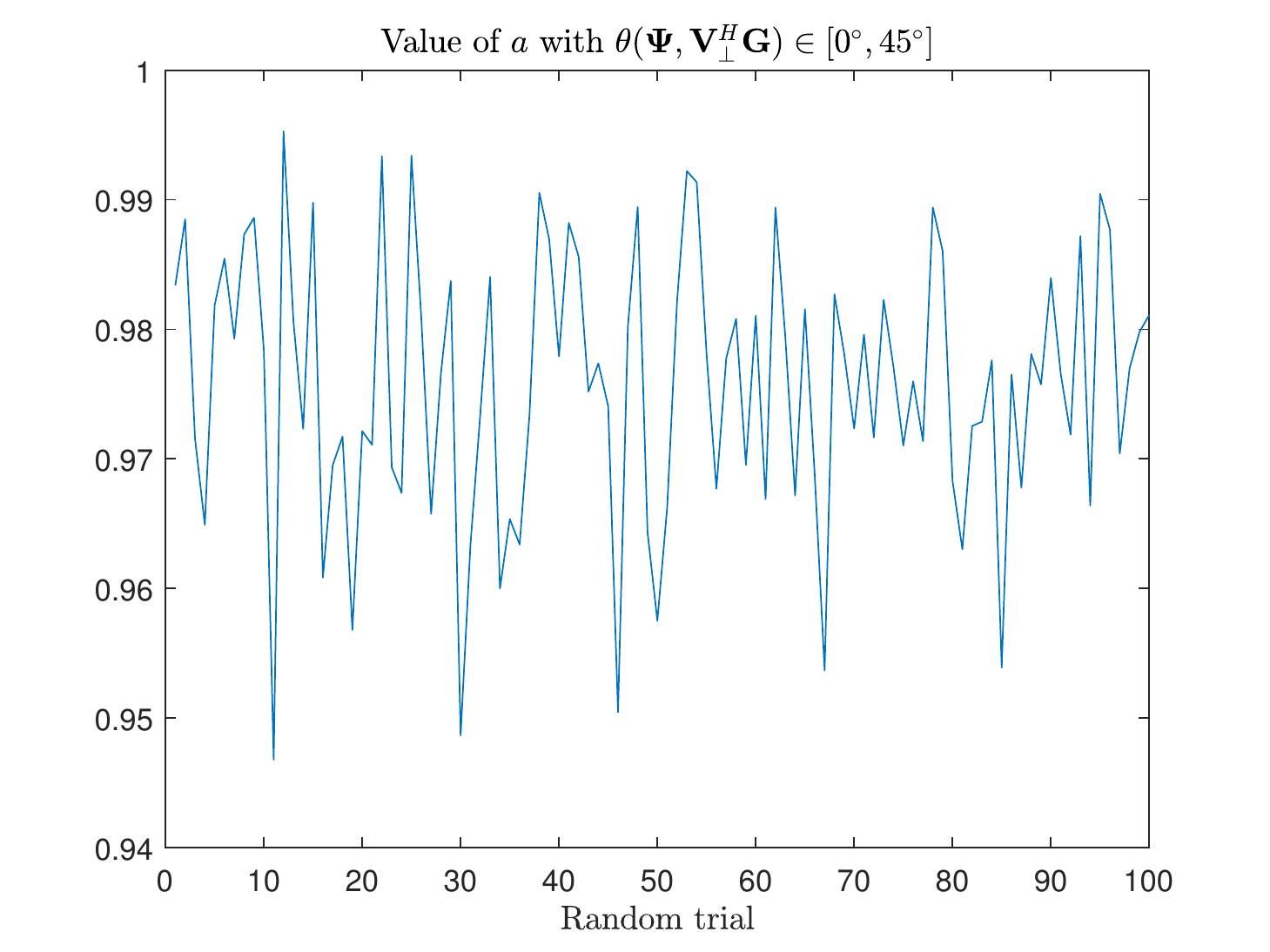}}
\caption{Value of $a$ when $\mPsi$ is picked at random or close to the principal subspace of $\C = \G\G^{H} + \gamma\eye{N}$.}
\label{fig:a_vs_Psi}
\end{figure}

Nevertheless in practice one does not know the value of $a$ a particular choice of $\T$ will result in. For illustration purposes let us assume that $\C = \G\G^{H} + \gamma\eye{N}$ and let us choose $\T = \begin{bmatrix} \isqnorm{\vv} \vv & \Vorth \mPsi \end{bmatrix}$. We consider 2 choices for $\mPsi$. In one case  $\mPsi$ is picked at random while in a second case $\mPsi$ is such that the angles between $\range{\Vorth^{H}\G}$ and $\range{\mPsi}$ are less than $45^{\circ}$. The corresponding values of $a$ are given in Figure \ref{fig:a_vs_Psi} for $100$ different trials of $\mPsi$. It is clear that picking $\mPsi$ at random does not offer much guarantee : some values of $a$ can be very small even though some can be rather high. Consequently selecting one random $\mPsi$ is risky, yet the idea of using a few random matrices $\mPsi$ is definitely not irrelevant, we will come back later on this point. On the contrary it is seen that if $\mPsi$ is selected such that $\range{\Vorth^{H}\G}$ and $\range{\mPsi}$ are close, $a$ is very close to $1$ which, following Figure \ref{fig:pdf_sinrloss_PA-MVDR}, should result in a very effective partially adaptive processor. This is a main argument for partially adaptive processors based on using the principal subspace of $\St$.

For the sake of completeness, let us consider the MPDR scenario where $\Ct=\gamma\C+P\vv\vv^{H}$. In this case, $\mOmegatilde_{22}=\gamma^{-1}\eye{R}$ -which implies that $\tilde{\lambda}_{i}=\gamma^{-1}$- and $\mOmegatilde_{21}=\vzero$ which leads to  $\tilde{\delta}_{i}=0$. Moreover one can show that
\begin{equation}
\frac{\vvtilde^{H}\Ctilde^{-1}\vvtilde}{\vvtilde^{H}\Cttilde^{-1}\vvtilde} = \gamma \left(1+P\gamma^{-1}\vvtilde^{H}\Ctilde^{-1}\vvtilde\right)	
\end{equation}
Finally we end up with the statistical representation in the MPDR scenario:
\begin{equation}\label{rep_SNRloss_PA-MPDR}
\losspampdr \dist a \left[1+(1+a\gamma^{-1}\SNRopt) \frac{\Cchisquare{R}{0}}{\Cchisquare{K-R+1}{0}}\right]^{-1}
\end{equation}
whose distribution can be written as
\begin{equation}\label{pdf_SNRloss_PA-MPDR}
\pdflosspampdr(\loss|a) = \pdflosspamvdr(\rho|a) \times \frac{\left[1+a\gamma^{-1}\SNRopt\right]^{K-R+1}}{\left[1+\loss\gamma^{-1}\SNRopt\right]^{K+1}}
\end{equation}
It is of interest to note the similarity between  equation \eqref{pdf_SNRloss_PA-MPDR} and equation \eqref{pdf_SNRloss_MPDR} which relates the SNR loss in the MPDR and the MVDR scenarios for a fully adaptive processor. Setting $R=N-1$ -which yields $a=1$- in \eqref{pdf_SNRloss_PA-MPDR} one recovers \eqref{pdf_SNRloss_MPDR}.

\subsection{Principal component based partially adaptive processing}
\begin{figure}[h]
\centering
\includegraphics[width=11cm]{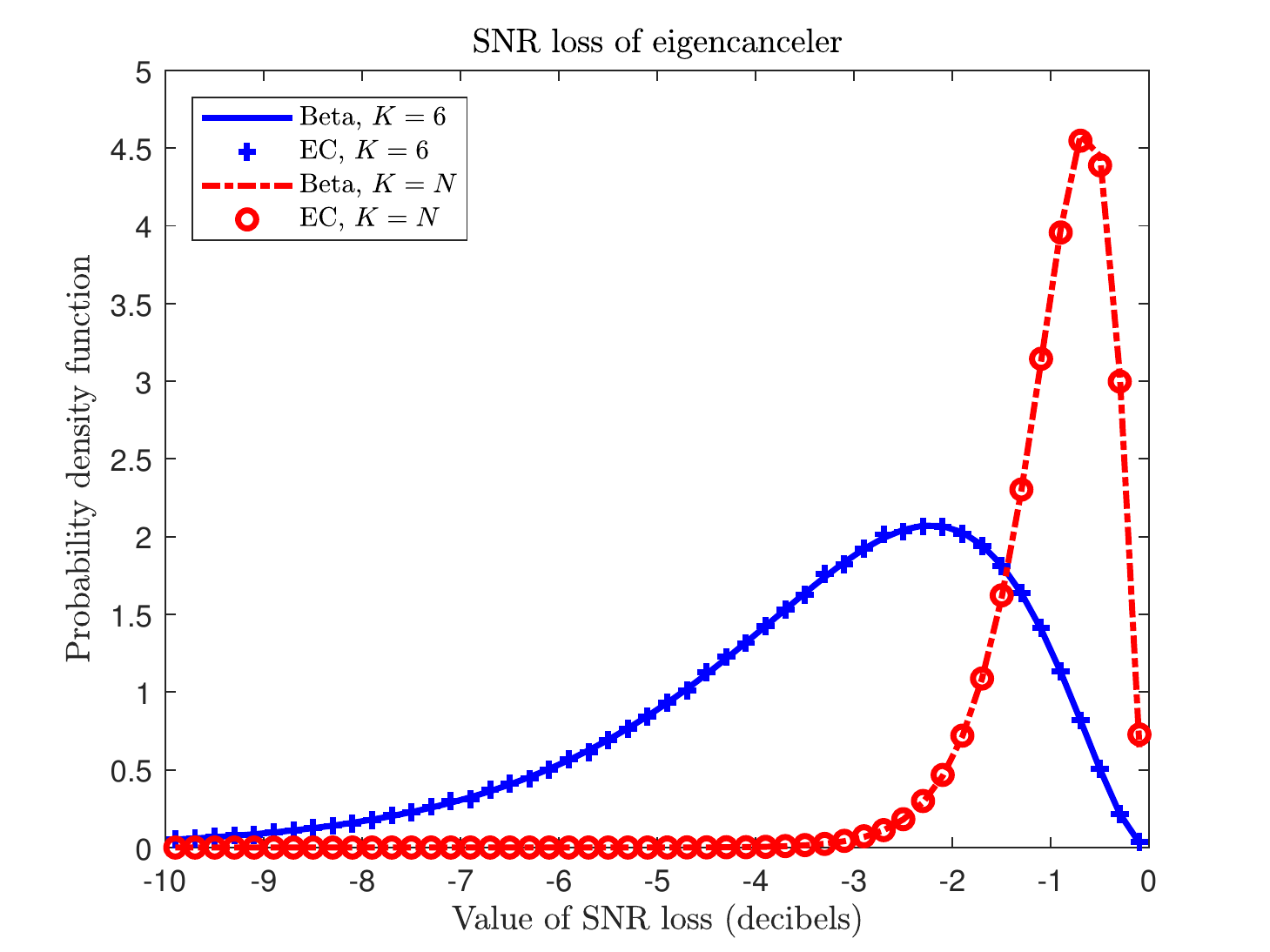} 
\caption{Probability density function of $\lossec$ and comparison with its $\betapdf{R}{K-R+1}$ prediction. $N=16$ and $R=3$.}
\label{fig:pdf_snrloss_ec_vs_K_R=3}
\end{figure}

As shown earlier, partially adaptive processing is particularly suitable when the noise covariance matrix is the sum of a strong low-rank component and white noise. In this case the optimal filter indeed belongs to a subspace, see \eqref{wopt_lowrank+white}, and can be approximately written -see \eqref{wopt_approx__lowrank+white}- as a projection onto the orthogonal complement of the principal subspace of $\C$. Moreover there would be no loss if one would be able to select $\T = \begin{bmatrix} \vv  & \G \end{bmatrix}$ or equivalently to select $\mPsi = \Vorth^{H}\G$. These observations have led to the use of what we will refer to as the class of \emph{principal component} adaptive processing whose gist is to use an estimate of $\range{\G}$ based on the principal eigenvectors of $\St$. There is a long list of references about such approach but the fundamentals are described in \cite{Kirsteins91,Haimovich91,Haimovich96}. In these references the so-called eigencanceler is defined as
\begin{equation}\label{eigencanceler}
\wec = \vv - \sum_{r=1}^{R} (\vu_{r}^{H}\vv	)\vu_{r}
\end{equation}
where $\vu_{r}$, $r=1\ldots N$ stands for the eigenvectors of $\St$ where the eigenvalues are arranged in descending order, i.e.
\begin{equation}\label{eig_St}
\St = 	\sum_{n=1}^{N} \lambda_{n} \vu_{n}\vu_{n}^{H}; \quad \lambda_{1} \geq \lambda_{2} \geq \cdots \geq \lambda_{N}
\end{equation}
The form in \eqref{eigencanceler} is in fact the sample version of \eqref{wopt_approx__lowrank+white}. To our knowledge there does not exist any exact analysis of the distribution of the SNR loss $\lossec$  associated with $\wec$. Only approximations of its distribution  are available under the assumption of $\C = \G\G^{H} + \gamma\eye{N}$ and $\G\G^{H} \gg \gamma\eye{N}$. To be more precise, Kirsteins and Tufts \cite{Kirsteins91} showed that, under the latter assumption,
\begin{equation}\label{rep_SNRloss_EC}
\boxed{\lossec \approxdist 	 \betapdf{R}{K-R+1}}
\end{equation}
which, when looking at \eqref{rep_SNRloss_PA-MVDR}, would mean that the eigencanceler would achieve a value of $a=1$, i.e., the value obtained with the clairvoyant choice of $\T$. It may be felt as a rather optimistic approximation but can predict fairly well the actual distribution of $\lossec$ when $\C$ is the sum of a very powerful low-rank term and a scaled identity matrix. This is  illustrated in Figure \ref{fig:pdf_snrloss_ec_vs_K_R=3} where the actual distribution of $\lossec$ obtained from Monte-Carlo simulations is compared with the $\betapdf{R}{K-R+1}$ distribution.

Again we remind that the scenario of this simulation is extremely favourable to the eigencanceler. With a smoother variation of the eigenvalues of $\C$ this conclusion should be re-examined.

A few observations can be made concerning the above technique. First note that it uses the eigenvectors associated with the largest eigenvalues. Another meaningful approach was proposed by Goldstein and Reed \cite{Goldstein97,Goldstein97b,Goldstein97c} called the cross spectral metric (CSM) whose principle is to select the eigenvectors that contribute most to increasing the SNR. 

 Note that the original eigencanceler makes use of the eigenvalue decomposition to obtain an estimate of the principal subspace of $\C$. However this is not the only method for that purpose. A very interesting approach which avoids eigenvalue decomposition was proposed by Goldstein, Reed and Scharf in \cite{Goldstein98} and referred to as the multistage Wiener filter (MWF), see also \cite{Honig01,Honig02}. It turns out that, at each stage, the MWF operates in the Krylov subspace of the sample covariance matrix, as does the conjugate gradient method at each of its iterations to solve a linear system of the type $\St \vw = \vv$. This analogy puts \textit{conjugate gradient} as an effective and efficient reduced-rank method for partially adaptive processing. 
 
 In \cite{Besson20c} we proposed to use a \textit{partial Cholesky factorization} as a way to approximate the principal subspace.  The partial Cholesky factor of $\C$ of rank $R$ will be denoted as $\pchol{\C}{R}$. It is  the $N|R$ (with $R \leq \rank{\C}$) lower triangular matrix with positive diagonal elements $\G_{\C} = \begin{pmatrix}  \G_{\C1}  \\ \G_{\C2} \end{pmatrix}$, where $\G_{\C1}$ is a $R \times R$ lower triangular matrix with positive diagonal elements and $\G_{\C2}$ is a $(N-R)|R$ matrix, defined from
 \begin{align}\label{def_pchol}
\C &= \begin{pmatrix} \C_{11} & \C_{12} \\ \C_{21} & \C_{22} \end{pmatrix} \nonumber \\
&= \begin{pmatrix} \C_{11} & \C_{12} \\ \C_{21} & \C_{21} \C_{11}^{-1} \C_{12} \end{pmatrix} + \begin{pmatrix} \mat{0} & \mat{0} \\ \mat{0} & \C_{2.1} \end{pmatrix} \nonumber \\
&= \begin{pmatrix}  \G_{\C1}  \\ \G_{\C2} \end{pmatrix}  \begin{pmatrix} \G_{\C1}^{H}  & \G_{\C2}^{H} \end{pmatrix}+ \begin{pmatrix} \mat{0} & \mat{0} \\ \mat{0} & \C_{2.1} \end{pmatrix}
 \end{align} 
 where $\C_{2.1} = \C_{22} - \C_{21} \C_{11}^{-1} \C_{12}$, $\G_{\C1}\G_{\C1}^{H} = \C_{11}$ and $\G_{\C2}\G_{\C1}^{H} = \C_{21}$. From a practical point of view, $\G_{\C}$ can be obtained, e.g., by using only $R$ steps of Algorithm 4.2.2 of \cite{Golub96}. In practice we can use as an alternative to \eqref{eigencanceler} the vector
\begin{equation}\label{w_pchol}
\wecpchol = \Projorth{\pchol{\St}{R}} 	\vv
\end{equation}
 An analysis was conducted in \cite{Besson20c} showing that
 \begin{equation}\label{rep_SNRloss_PCHOL}
\lossecpchol \approxdist 	a' \, \betapdf{R}{K-R+1}
 \end{equation}
where $a'$ is a scalar that depends on $\C$ and is rather close to $1$. The representation in \eqref{rep_SNRloss_PCHOL} is compliant with that of \eqref{rep_SNRloss_PA-MVDR} and predicts that the partial Cholesky factorization could perform well.  Actually it was shown that the distributions of $\lossec$ and $\lossecpchol$ are almost identical, with the latter possibly better when $\Ct=\C+P\vv\vv^{H}$. Therefore, the computationally simpler partial Cholesky factorization is a very good alternative to eigenvalue decomposition.

Finally we would like to draw attention to the fact that all these partially adaptive processors require the selection of $R$. Obviously, when $\C=\G\G^{H}+\gamma\eye{N}$ and the first low-rank term is predominant, $R$ should be chosen as the rank of $\G$. Actually selecting $R$ below this rank has dramatic consequences with very low SNR, mainly due to the fact that the range space of $\G$ cannot be captured by $\T$. Over-estimating $R$ is less consequential, yet it leads to performance loss. 

\subsection{Partially adaptive processing with random transformations}
\begin{figure}[h]
\centering
\includegraphics[width=14cm]{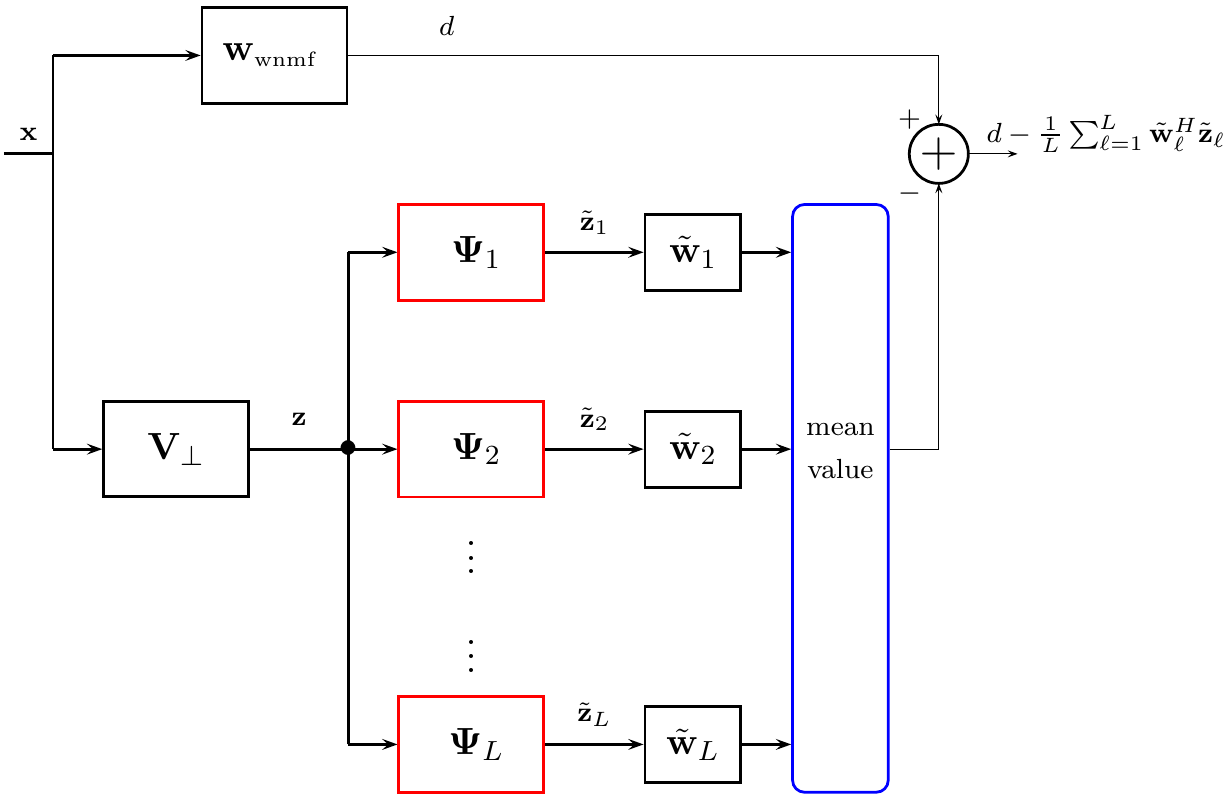}
\caption{Structure of Marzetta \emph{et al.} partially adaptive processor based on random reduced-dimension transformations.}
\label{fig:Marzetta}
\end{figure}

We make a digression here and consider an interesting class of partially adaptive processors based on random $\T$, actually random $\mPsi$. As said before, if one picks $\mPsi$ at random then the coefficient $a = \frac{\vvtilde^{H}\Ctilde^{-1}\vvtilde}{\vv^{H}\C^{-1}\vv} $ which intervenes in \eqref{rep_SNRloss_PA-MVDR} can sometimes be very close to $1$ (hence a good filter) yet other times small. This is undesirable in practice as one cannot control the performance of the partially adaptive processor which depends on the particular outcome of $\mPsi$. However using and improving over this basic idea results in a rather efficient scheme as proposed in \cite{Marzetta11}. The idea of Marzetta \emph{et al.} is to use a certain number or random matrices $\mPsi_{\ell}$ and to average over the corresponding filters, as illustrated in Figure \ref{fig:Marzetta}.

\begin{figure}[h]
	\centering
	\includegraphics[width=11cm]{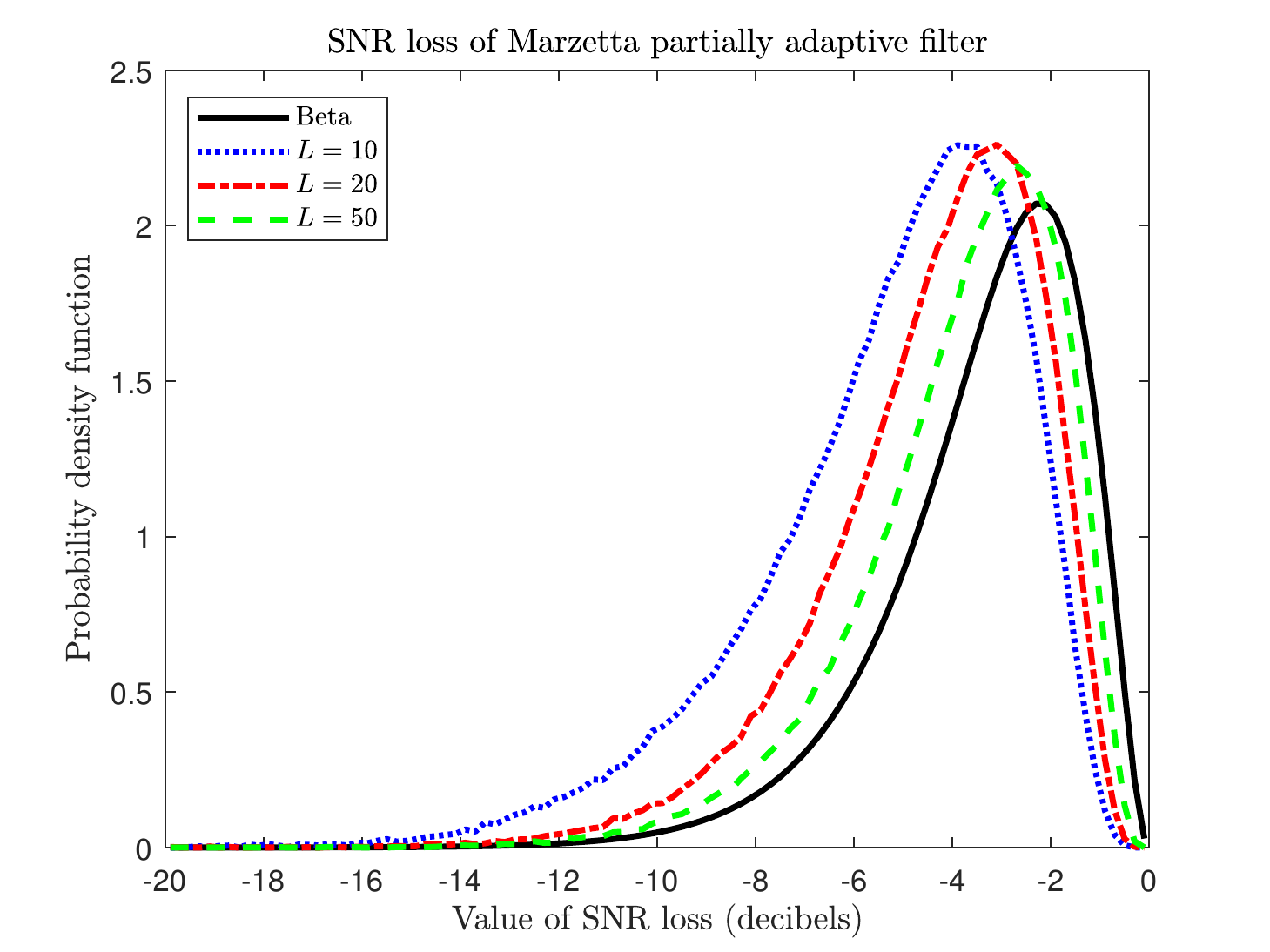}
	\caption{Probability density function of Marzetta's partially adaptive scheme SNR loss. $N=16$ $K=6$ and $R=3$.}
	\label{fig:pdf_snrloss_marzetta_R=3}
\end{figure}

\begin{figure}[h]
	\centering
	\includegraphics[width=11cm]{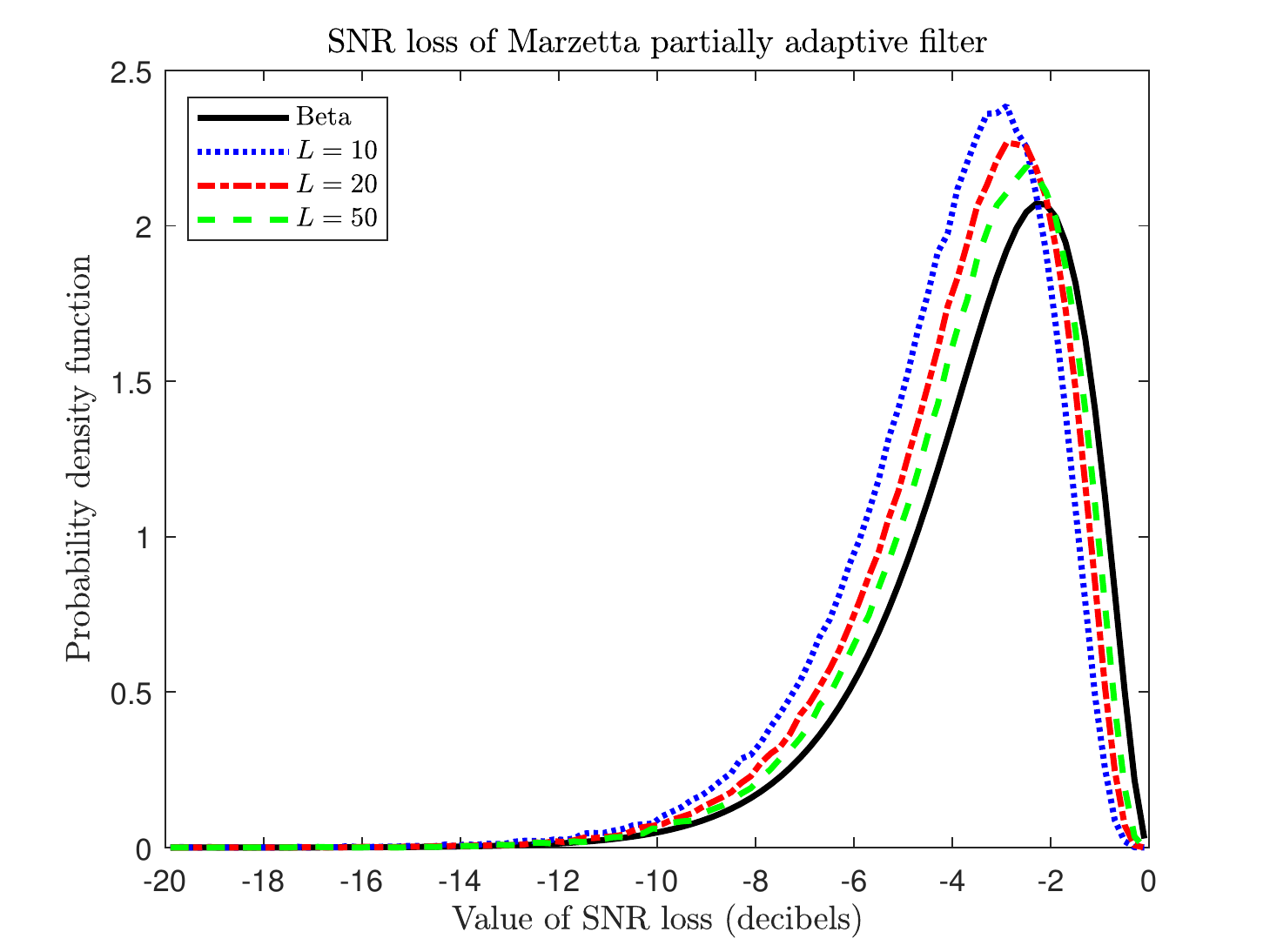}
	\caption{Probability density function of Marzetta's partially adaptive scheme SNR loss. $N=16$, $K=6$ and $R=4$.}
	\label{fig:pdf_snrloss_marzetta_R=4}
\end{figure}

The final adaptive filter writes
\begin{align}\label{w_Marzetta}
\wmarzetta = \wcbf - \frac{1}{L}\sum_{\ell=1}^{L} \Vorth \mPsi_{\ell} (\mPsi_{\ell}^{H}\Vorth^{H}\St\Vorth\mPsi_{\ell})^{-1}\mPsi_{\ell}^{H}\Vorth^{H}\St\wcbf
\end{align}
In \cite{Marzetta11} it is proposed to draw the matrices $\mPsi_{\ell}$ from a uniform distribution on the Stiefel manifold of $(N-1)|R$ semi-unitary matrices. However the vector in \eqref{w_Marzetta} remains the same if $\mPsi_{\ell}$ are drawn from a $\mCN{N-1,R}{\Mzero}{\eye{N}}{\eye{R}}$. Analysis of the SNR loss associated with $\wmarzetta$ is not available but very good performance was observed. Moreover a number of interesting results and insights are provided in \cite{Marzetta11} regarding the average value of $\mPsi(\mPsi^{H}\mOmega\mPsi)^{-1}\mPsi^{H}$ for arbitrary positive semi-definite matrices $\mOmega$.  For illustration purposes, we display in Figure \ref{fig:pdf_snrloss_marzetta_R=3} the SNR loss of $\wmarzetta$ when $R=3$, i.e., when $R$ corresponds to the number of interfering signals.
As can be seen one does not reach the performance of the eigencanceler predicted by \eqref{rep_SNRloss_EC}  but the difference is small. However we observe that improvement can be achieved when $R$ is slightly above the number of interfering signals, as depicted in Figure \ref{fig:pdf_snrloss_marzetta_R=4}. Therefore one advantage of this method is that one not needs to know precisely the rank of the true interference covariance matrix.
\subsection{Synthesis}
\begin{synthesis}
In this section we looked at partially adaptive processing, i.e., when the adaptive filter belongs to a given subspace of the entire observation space. We begun with a fixed reduced-dimension transformation $\T$ and a known $\C$ and we showed that this kind of reduced-dimension structure can be as efficient as  the optimal processor when $\C$ contains a strong low-rank component and $\T$ is suitably chosen. In practice $\C$ is substituted for the sample covariance matrix and we showed that the SNR loss is a scaled beta distributed random variable for fixed $\T$. Next we addressed adaptive techniques where $\T$ is selected from the data with the main purpose of retrieving the principal subspace of $\C$. In this case no exact distribution of the SNR loss is available but a rather accurate beta distribution approximation exists which shows a better convergence of partially adaptive filters compared to  fully adaptive filters. Finally we hinted at an original idea developed by Marzetta \emph{et al.} relying on random reduced-dimension transformations.
\end{synthesis}
\section{Regularized adaptive processing}
One of the most widely used method to cope with the effects of small sample support on conventional adaptive filters is to use regularization and to solve
\begin{equation}\label{pb_DL}
\underset{\vw}{\min}	\underset{\text{output power}}{\vw^{H}\Chat\vw} + \underset{\text{regularization}}{\mu \sqnorm{\vw}} \text{ subject to } \underset{\text{constraint}}{\vw^{H}\vv=1}
\end{equation}
where $\Chat= K^{-1}\St$. In adaptive beamforming the penalizing term $\mu \sqnorm{\vw}$ is inversely proportional to the white noise array gain and hence fills the purpose of ensuring that the latter is not too low. The solution is given by
\begin{equation}\label{w_DL}
\wdl  = \frac{(\St+K \mu \eye{N})^{-1}\vv}{\vv^{H}(\St+K\mu\eye{N})^{-1}\vv}	
\end{equation}
and is often referred to as \textbf{\emph{diagonal loading}} \cite{Carlson88}. Diagonal loading is an ubiquitous technique that appears as the solution of many problem formulations in robust adaptive beamforming, see e.g., \cite{Cox87,Li03,Li03b,Vorobyov03,Gershman03,Shahbazpanahi03}. Although diagonal loading has been dealt with in hundreds of papers to the best of our knowledge there is no exact analysis of the SNR loss associated with $\wdl$ for arbitrary $\C$. Only approximations are available in the case $\C = \G\G^{H} + \gamma\eye{N}$ and $\G\G^{H} \gg \gamma\eye{N}$. These approximations require that $\mu$ is larger than $\gamma$ (i.e., the loading level  is larger than the white noise power), yet much below the eigenvalues of $\G\G^{H}$, which is possible only if a very large gap exists between the latter and the white noise power. Moreover, they also assume that $K$ is small and typically slightly above the rank of $\G$. The fundamental works where derivation and analysis of this technique can be found are \cite{Abramovich81,Cheremisin82}. In the sequel we provide a sketch of the derivations leading to the distribution of the SNR loss of $\wdl$ when $\C = \G\G^{H} + \gamma\eye{N}$. Let us write the eigenvalue decomposition of $\C$ as
\begin{align}\label{eig_C}
\C &= \G\G^{H} + \gamma\eye{N} \nonumber \\
&= \begin{bmatrix} \U_{s} & \U_{n} \end{bmatrix} \begin{bmatrix} \mLambda_{s} & \Mzero \\ \Mzero & \Mzero \end{bmatrix}	\begin{bmatrix} \U_{s}^{H} \\ \U_{n}^{H} \end{bmatrix} + \gamma\eye{N} \nonumber \\ 
&= \begin{bmatrix} \U_{s} & \U_{n} \end{bmatrix} \begin{bmatrix} \mLambda_{s}+\gamma\eye{R} & \Mzero \\ \Mzero & \gamma\eye{N-R} \end{bmatrix}	\begin{bmatrix} \U_{s}^{H} \\ \U_{n}^{H} \end{bmatrix} \nonumber \\
&=\U \mLambda \U^{H}
\end{align}
and let us assume that $\mLambda_{s}(r,r)  \gg \gamma$. Under this hypothesis, one has
\begin{equation}\label{approx_inv_Sigma}
\C^{-1} = \U_{s}(\mLambda_{s}+\gamma\eye{R})^{-1} \U_{s}^{H} + 	\gamma^{-1} \U_{n}\U_{n}^{H} \simeq \gamma^{-1} \U_{n}\U_{n}^{H} 
\end{equation}
which means that inverting $\C$ amounts to projecting onto the noise subspace. This is basically the main property that will be used below.

We begin with  the exact representation of $\St+K\mu\eye{N}$ as
\begin{align}
\St+K\mu\eye{N} &\dist \U \mLambda^ {1/2} \N \N^{H}	\mLambda^ {1/2} \U^{H} +K\mu\eye{N} \nonumber \\
&= \U \mLambda^ {1/2} \left(\N \N^{H} + K\mu \mLambda^{-1}\right) \mLambda^ {1/2} \U^{H} \nonumber \\
&= \U \mLambda^ {1/2} \Wtilde \mLambda^ {1/2} \U^{H}
\end{align}
where $\N \dist \mCN{N,K}{\Mzero}{\eye{N}}{\eye{K}}$ and  $\Wtilde = \N \N^{H} + K\mu \mLambda^{-1}$.
It follows that
\begin{align}\label{deriv_rep_rho_DL}
\lossdl &= \frac{[\vv^H (\St+K\mu\eye{N})^{-1} \vv]^{2}}{(\vv^{H}\C^{-1}\vv)[\vv^{H}(\St+K\mu\eye{N})^{-1}\C(\St+K\mu\eye{N})^{-1}\vv]} \nonumber \\
&\dist  \frac{[\vv^H \U \mLambda^ {-1/2} \Wtilde^{-1} \mLambda^ {-1/2} \U^{H}  \vv]^{2}}{(\vv^{H}\C^{-1}\vv)[\vv^{H}\U \mLambda^ {-1/2}  \Wtilde^{-1}  \mLambda^ {-1/2} \U^{H} \U \mLambda \U^{H} \mLambda^ {-1/2} \U^{H} \Wtilde^{-1}  \mLambda^ {-1/2} \U^{H} \vv]} \nonumber \\
&\dist \frac{[\vv^H \U \mLambda^ {-1/2} \Wtilde^{-1} \mLambda^ {-1/2} \U^{H}  \vv]^{2}}{(\vv^{H}\C^{-1}\vv)[\vv^{H}\U \mLambda^ {-1/2}  \Wtilde^{-2}  \mLambda^ {-1/2} \U^{H} \vv]} 
\end{align}
Now we use the fact that $\mLambda_{s}(r,r) \gg \gamma$ to write 
\begin{align}
\Wtilde &= \N \N^{H} + K\mu \mLambda^{-1} = \begin{bmatrix} \N_{s} \\ \N_{n} \end{bmatrix} \begin{bmatrix} \N_{s}^{H} & \N_{n}^{H} \end{bmatrix}+ K\mu \begin{bmatrix} (\mLambda_{s}+\gamma\eye{R})^{-1} & \Mzero \\ \Mzero & \gamma^{-1}\eye{N-R} \end{bmatrix} \nonumber \\
&\simeq  \begin{bmatrix} \N_{s}\N_{s}^{H} & \N_{s}\N_{n}^{H} \\ \N_{n}\N_{s}^{H} & \N_{n}\N_{n}^{H} + K \mu\gamma^{-1}\eye{N-R} \end{bmatrix}  \nonumber \\
&=  \begin{bmatrix} \Wtilde_{ss} & \Wtilde_{sn} \\ \Wtilde_{ns} & \Wtilde_{nn}  \end{bmatrix}  \nonumber \\
\end{align}
and
\begin{align}
\mLambda^ {-1/2} \U^{H} \vv \simeq \begin{bmatrix} \vzero \\ \gamma^ {-1/2} \U_{n}^{H} \vv\end{bmatrix} = \gamma^ {-1/2} \begin{bmatrix}\vzero \\ \vvn \end{bmatrix}
\end{align}
These approximations result in
\begin{align}
\lossdl &\simeq  \frac{\left(\begin{bmatrix} \vzero & \vvn^H \end{bmatrix} \Wtilde^{-1}  \begin{bmatrix} \vzero \\ \vvn \end{bmatrix} \right)^{2}}{(\vvn^{H}\vvn) \left(\begin{bmatrix} \vzero & \vvn^H \end{bmatrix} \Wtilde^{-2}\begin{bmatrix} \vzero \\ \vvn \end{bmatrix}\right)} \nonumber \\
&= \frac{[\vvn^H \Wtilde_{n.s}^{-1}  \vvn]^{2}}{(\vvn^{H}\vvn)[\vvn^{H}\Wtilde_{n.s}^{-1}(\eye{N-R}+\Wtilde_{ns}\Wtilde_{ss}^{-2}\Wtilde_{sn})\Wtilde_{n.s}^{-1}\vvn]} 
\end{align}
At this stage further approximations are required to obtain tractable expressions and eventually arrive at the final representation in \eqref{rep_SNRloss_DL}. In \cite{Cheremisin82} two successive approximations are actually made while \cite{Abramovich00} directly states the expression \eqref{rep_SNRloss_DL_approx} below. Whatever, it amounts to use the following approximations:
\begin{align}\label{approx_Wtilde_n.s}
\Wtilde_{n.s}^{-1} &= [\Wtilde_{nn}-\Wtilde_{ns}\Wtilde_{ss}^{-1}\Wtilde_{sn}] ^{-1} \nonumber \\
&\simeq [K \mu\gamma^{-1}\eye{N-R} + \W_{nn}-\W_{ns}\W_{ss}^{-1}\W_{sn}] ^{-1} \nonumber \\
&=[K\mu\gamma^{-1}\eye{N-R} + \N_{n} \left[\eye{K}-\N_{s}^{H}(\N_{s}\N_{s}^{H})^{-1}\N_{s}\right]\N_{n}^{H}]^{-1} \nonumber \\
&=[K\mu\gamma^{-1}\eye{N-R} + \N_{n} \Projorth{\N_{s}^{H}}\N_{n}^{H}]^{-1} \nonumber \\
&=[K\mu\gamma^{-1}\eye{N-R} + \N_{n} \H_{s} \H_{s}^{H}\N_{n}^{H}]^{-1} \nonumber \\
&=(K\mu\gamma^{-1})^{-1} [\eye{N-R}-\N_{n} \H_{s} (K\mu\gamma^{-1}\eye{R}+\H_{s}^{H}\N_{n}^{H}\N_{n} \H_{s})^{-1}\H_{s}^{H}\N_{n}^{H}] \nonumber \\
&\simeq (K\mu\gamma^{-1})^{-1} [\eye{N-R}-(K\mu\gamma^{-1})^{-1}\N_{n} \H_{s} \H_{s}^{H}\N_{n}^{H}] \nonumber \\
&\simeq(K\mu\gamma^{-1})^{-1} \eye{N-R}
\end{align}
and
\begin{align}
\Wtilde_{ns}\Wtilde_{ss}^{-2}\Wtilde_{ns} &\simeq \W_{ns}\W_{ss}^{-2}\W_{sn} \nonumber \\
&= \left[\N_{n}\N_{s}^{H} (\N_{s}^{H}\N_{s})^{-\onehalf}\right] \W_{ss}^{-1} \left[\N_{n}\N_{s}^{H} (\N_{s}^{H}\N_{s})^{-\onehalf}\right]^{H}
\end{align}

\begin{figure}[h]
	\centering
	\includegraphics[width=11cm]{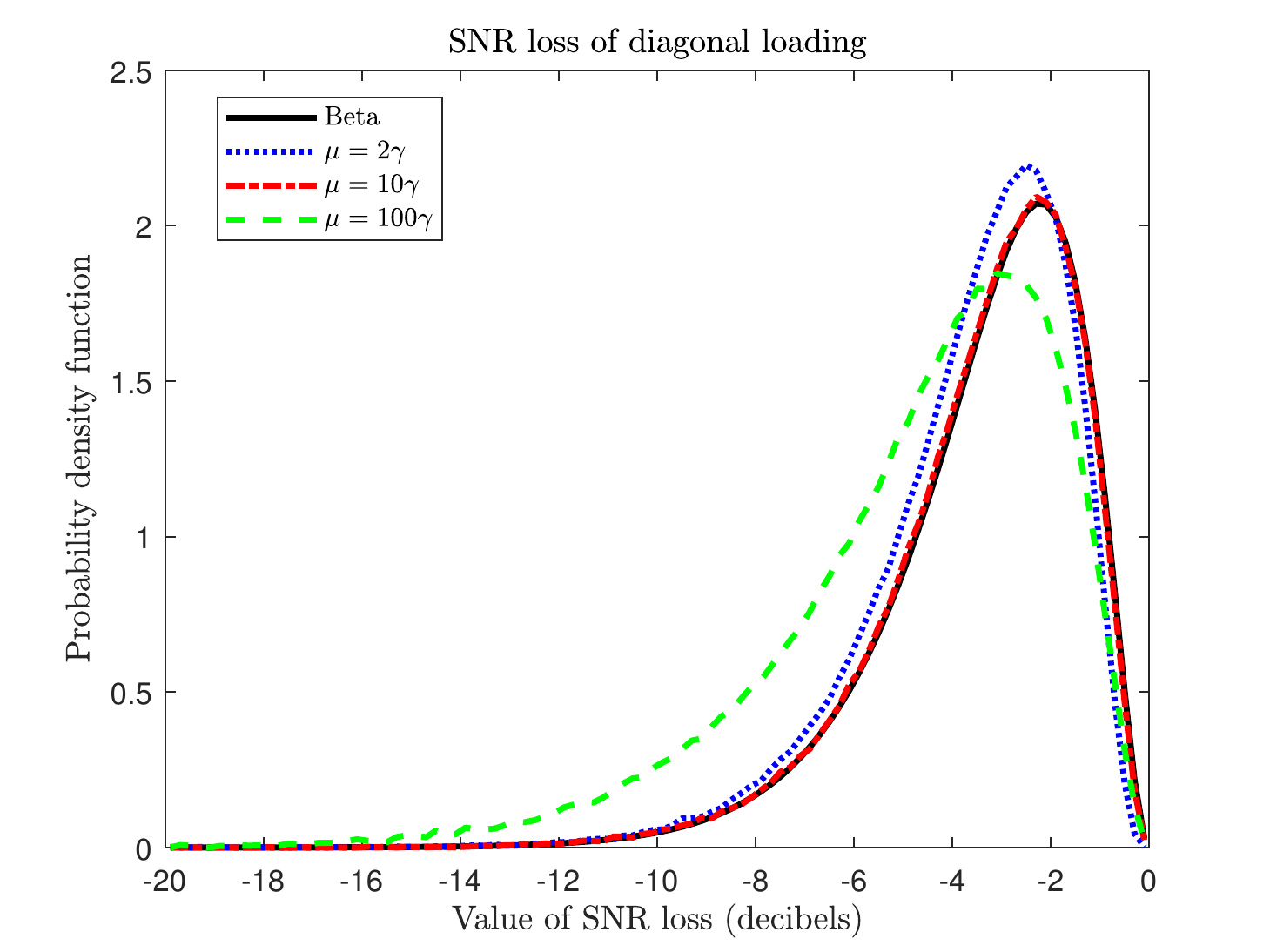}
	\caption{Probability density function of the SNR loss with diagonal loading. $N=16$ and $K=6$.}
	\label{fig:pdf_snrloss_dl}
\end{figure}

Note that $\N'_{n}=\N_{n}\N_{s}^{H} (\N_{s}^{H}\N_{s})^{-\onehalf}$ is independent of $\N_{s}$ (hence of $\W_{ss}$) and follows a $\mCN{N-R,R}{\Mzero}{\eye{N-R}}{\eye{R}}$ distribution while $\W_{ss}=\N_{s}\N_{s}^{H} \dist \CW{R}{K}{\eye{R}}$. 
Therefore
\begin{align}\label{rep_SNRloss_DL_approx}
\lossdl \simeq 	\frac{(\vvn^H \vvn)^{2}}{(\vvn^{H}\vvn)[\vvn^{H}\left(\eye{N-R}+\N'_{n}\W_{ss}^{-1}(\N'_{n})^{H}\right)\vvn]}  
\end{align}
From $\W_{ss}=\N_{s}\N_{s}^{H} \dist \CW{R}{K}{\eye{R}}$ and $\N'_{n} \dist \mCN{N-R,R}{\Mzero}{\eye{N-R}}{\eye{R}}$, we have that
\begin{align}
\vvn^{H}\N'_{n}\W_{ss}^{-1}(\N'_{n})^{H}\vvn & \dist (\vvn^{H}\N'_{n}(\N'_{n})^{H}\vvn) / \Cchisquare{K-R+1}{0} \nonumber \\
&\dist (\vvn^{H}\vvn) \Cchisquare{R}{0} / \Cchisquare{K-R+1}{0}
\end{align} 
It yields
\begin{equation}\label{rep_SNRloss_DL}
\boxed{\lossdl \approxdist \left[1 + \frac{\Cchisquare{R}{0}}{\Cchisquare{K-R+1}{0}}\right]^{-1}	\dist \betapdf{R}{K-R+1}}
\end{equation}
which coincides with the distribution of the SNR loss of the eigencanceler in \eqref{rep_SNRloss_EC}. This is actually not surprising. Indeed, if the eigenvalue decomposition of $\St$ writes as in \eqref{eig_St} with $\lambda_{R} \gg \lambda_{R+1} \simeq \lambda_{R+2} \simeq \cdots \simeq \lambda_{N}$ then
\begin{align}\label{DL_EC}
(\St+K\mu\eye{N})^{-1}\vv &= \left(\sum_{n=1}^{N}(\lambda_{n}+K\mu)^{-1}\vu_{n}\vu_{n}^{H}\right)	\vv \nonumber \\
&\simeq  \left(\sum_{n=R+1}^{N}(\lambda_{n}+K\mu)^{-1}\vu_{n}\vu_{n}^{H}\right)	\vv \nonumber \\
&\simeq \eta \left(\sum_{n=R+1}^{N}\vu_{n}\vu_{n}^{H}\right)	\vv \nonumber \\
&= \eta \left(\eye{N} - \sum_{r=1}^{R}\vu_{r}\vu_{r}^{H}\right)	\vv \nonumber \\
&\propto \wec
\end{align}
and therefore diagonal loading more or less behaves like the eigencanceler whose SNR loss distribution is given by \eqref{rep_SNRloss_EC}. Numerical simulations tend to confirm this fact and show that diagonal loading is very efficient in low sample support with strong low-rank interference, provided that the loading level $\mu$ is chosen properly. This is illustrated in Figure \ref{fig:pdf_snrloss_dl} where, if $\mu$ is chosen properly, the filter $\wdl$ has a  performance commensurate with that of the eigencanceler. Note also that when $\mu$ is not sufficiently larger than $\gamma$ or sufficiently smaller than the largest eigenvalues then the approximation does not hold.

As said before exact analysis of the SNR loss of $\wdl$ does not exist, the main technical problem being that the Wishart distribution of $\St$ is lost when adding the scaled identity matrix. There have been attempts at asymptotic (i.e., when $K \rightarrow \infty$) analysis, see \cite{Ganz90,Dilsavor93,Fertig00}. While these analyses are interesting they hold under a framework (large $K$) which does not correspond to the framework of most interest of diagonal loading, namely low sample support. We believe that a much more adequate and solid framework is that of \textbf{\textit{random matrix theory}} (RMT) for which $N,K \rightarrow \infty$ with $N/K \rightarrow c$. The beauty of RMT lies in 1)a very rigorous theory which applies to large class of processes and 2)its ability to predict fairly well what happens in finite samples despite its ``asymptotic'' nature, certainly because $K$ scales with $N$. Among the numerous important works in this area we would like to excerpt papers by Mestre and his co-authors \cite{Mestre06,Rubio12} which deal with diagonal loading. In these references a central limit theorem is provided showing that $b_{N}\sqrt{K}\left(\SNR(\wdl)-\overline{\SNR}(\wdl)\right)$ converges to a normalized Gaussian distribution $\mathcal{N}(0,1)$.

\section{Conclusion}
In this report we provided a short overview of many works spanned over fifty years to analyze the distribution of the SNR loss at the output of multichannel adaptive filters trained with a finite number of snapshots. We successively investigated fully adaptive processing, partially adaptive processing and regularized adaptive processing and tried to summarize the main results. We also tried to show  how powerful are the statistical tools related to matrix-variate distributions in deriving these results.  As said before this presentation is not exhaustive and probably a lot of interesting results are missing. In particular we only alluded to RMT which, in the recent years, has proved to provide tremendous tools to analyze the performance of many array processing methods. 
\appendix
\section{Complex matrix-variate distributions}
In this appendix, we briefly give the definitions and some properties of the complex matrix-variate distributions used in this report. 
\subsection*{Gaussian distribution}
The complex matrix-variate Gaussian distribution is denoted by $\mCN{N,K}{\Xbar}{\mSigma}{\mOmega}$ with a p.d.f. given by
\begin{equation}\label{pdf_Gaussian}
p(\X) = \pi^{-NK} \det{\mSigma}^{-K} \det{\mOmega}^{-N} \etr{-\mSigma^{-1}(\X-\Xbar)\mOmega^{-1}(\X-\Xbar)^{H}}	 \tag{Gaussian}
\end{equation}
For $K=1$ we note $\vCN{N}{\vxbar}{\mSigma}$ the distribution of the vector $\vx$.
\subsection*{Wishart distribution}
When $\X \dist \mCN{N,K}{\Mzero}{\mSigma}{\eye{K}}$ with $K \geq N$ then $\S=\X\X^{H} \dist \CW{N}{K}{\mSigma}$ follows a complex Wishart distribution with p.d.f. 
\begin{equation}\label{pdf_Wishart}
p(\S) \propto \det{\mSigma}^{-K} \det{\S}^{K-N} \etr{-\mSigma^{-1}\S} \tag{Wishart}
\end{equation}
where $\propto$ means ``proportional to''. Properties of partitioned Wishart matrices are very important to derive the SNR loss distribution and we briefly recall some of them. Let us partiton $\S$ as
\begin{equation}\label{partition_S}
\mat{S} = \begin{pmatrix} \underset{P,P}{\mat{S}_{11}} &
\underset{P,Q}{\mat{S}_{12}} \\
\underset{Q,P}{\mat{S}_{21}} & \underset{Q,Q}{\mat{S}_{22}}
\end{pmatrix}
\end{equation}
Then the inverse of $\mat{S}$ can be partitioned as
\begin{equation}
\S^{-1} = \begin{pmatrix} \S_{1.2}^{-1} & -\S_{1.2}^{-1}\S_{12}\S_{22}^{-1} \\
-\S_{2.1}^{-1}\S_{21}\S_{11}^{-1} & \S_{2.1}^{-1} \end{pmatrix}
\end{equation}
where $\S_{1.2}=\S_{11}-\S_{12}\S_{22}^{-1}\S_{21}$ and $\S_{2.1}=\S_{22}-\S_{21}\S_{11}^{-1}\S_{12}$. It has been shown that $\S_{1.2}$ and $\left(\S_{22}^{-1}\S_{21} , \S_{22}\right)$ are independently distributed with
\begin{align}
\S_{1.2} &\dist \CW{P}{K-Q}{\mat{\Sigma}_{1.2}} \\
\S_{22} &\dist \CW{Q}{K}{\mat{\Sigma}_{22}}
\end{align}
and the conditional distribution of the $Q|P$ matrix $\S_{22}^{-1}\S_{21}$ given $\S_{22}$ is
\begin{equation}
\S_{22}^{-1}\S_{21} | \S_{22} \dist \mCN{Q,P}{\mat{\Sigma}_{22}^{-1} \mat{\Sigma}_{21}}{\S_{22}^{-1}}{\mat{\Sigma}_{1.2}}.
\end{equation}
The marginal distribution of $\T_{21}=\S_{22}^{-1} \S_{12}$ is a matrix-variate Student distribution (given below), i.e., $\T_{21}  \dist \mCT{Q,P}{\mSigma_{22}^{-1} \mSigma_{21} }{K-Q+1}{\mSigma_{22}^{-1}}{ \mSigma_{1.2}}$.

The complex chi-square distribution with $N$ degrees of freedom and non-centrality parameter $\delta$ will be denoted as $\Cchisquare{N}{\delta}$.  It is the distribution of $\vx^{H}\vx$ when $\vx \dist \vCN{N}{\vxbar}{\eye{N}}$ and $\delta =  \vxbar^{H}\vxbar$.
\subsection*{Student distribution}
The complex matrix-variate $t$ distribution is denoted by $\mCT{N,K}{\nu}{\Xbar}{\mSigma}{\mOmega}$ and its p.d.f is given by 
\begin{equation}\label{pdf_matrix_Student}
p(\X) \propto \det{\mSigma}^{-K} \det{\mOmega}^{-N} \det{\I_{N}+\mSigma^{-1}(\X-\Xbar)\mOmega^{-1}(\X-\Xbar)^{H}}^{-(\nu+N+K-1)}	\tag{matrix-Student}
\end{equation}
It can be represented as 
\begin{align}
\X &\dist \Xbar + (\W_{1}^{-1/2})^{H}\Y_{1}; \quad &\W_{1} \dist \CW{N}{\nu+N-1}{\mSigma^{-1}}, \; \Y_{1} \dist \mCN{N,K}{\mat{0}}{\I_{N}}{\mOmega} \nonumber \\
\X &\dist \Xbar + \Y_{2}\W_{2}^{-1/2}; \quad &\Y _{2}\dist \mCN{N,K}{\mat{0}}{\mSigma}{\I_{K}}, \; \W \dist\CW{K}{\nu+K-1}{\mOmega^{-1}}
\end{align}
where $\Y_{i}$ is independent of $\W_{i}$.

For $K=1$ we let
\begin{equation}\label{pdf_vector_Student}
p(\vx) \propto \det{\mSigma}^{-1} \left[1+(\vx-\vxbar)^{H}\mSigma^{-1}(\vx-\vxbar)\right]^{-(\nu+N)}	\tag{vector-Student}
\end{equation}
and one has $\vx \dist \vxbar + \vCN{N}{\vzero}{\mSigma}/\sqrt{\Cchisquare{\nu}{0}}$. 
\subsection*{F distribution}
If $\S_{i} \dist \CW{N}{K_{i}}{\mSigma}$, $i=1,2$, then $\F=\S_{1}^{\onehalf} \S_{2}^{-1}\S_{1}^{\onehalf}$ -where $\S^{\onehalf}$ denotes the unique Hermitian square-root of $\S$- follows a complex matrix-variate $F$ distribution with p.d.f. 
 \begin{equation}\label{pdf_F}
p(\F) \propto \det{\F}^{K_{1}-N} \det{\I_{N}+\F}^{-(K_{1}+K_{2})} 	\tag{F}
 \end{equation}
and we note $\F \dist \CF{N}{K_{1}}{K_{2}}$. If $\F$ is partitioned as in \eqref{partition_S} then $\F_{1.2}$ and $\F_{22}$ are independent and
\begin{align}\label{pdf_F1.2_F22}
\F_{1.2}  \dist \CF{P}{K_{1}-Q}{K_{2}}; \quad \F_{22}  \dist \CF{Q}{K_{1}}{K_{2}-P}
\end{align}
\begin{equation}
\F_{22}^{-1}\F_{21} | 	\F_{1.2},\F_{22} \dist \mCT{Q,P}{K_{1}+K_{2}-N+1}{\Mzero}{\I_{Q}+\F_{22}^{-1}}{\I_{P} + \F_{1.2}}
\end{equation}
When $N=1$, $F \dist \Cchisquare{K_{1}}{0} / \Cchisquare{K_{2}}{0} \dist \CF{1}{K_{1}}{K_{2}}$.
\subsection*{Beta distribution}
Let $F \dist \CF{1}{K_{1}}{K_{2}}$. Then $B=(1+F)^{-1} \dist \betapdf{K_{1}}{K_{2}}$ is beta distributed and its probability density function is given by 
\begin{equation}\label{pdf_beta}
p(B) = \frac{1}{\Beta{K_{1}}{K_{2}}} \tag{Beta} B^{K_{2}-1}(1-B)^{K_{1}-1}
\end{equation}
where $\Beta{K_{1}}{K_{2}} = \frac{\Gamma(K_{1})\Gamma(K_{2})}{\Gamma(K_{1}+K_{2})}$.
%


\end{document}